 \documentclass[prb,aps,twocolumn,showpacs]{revtex4} 
\usepackage{graphicx} 
\usepackage{tabularx}
\usepackage{dcolumn} 
\usepackage{color}
\usepackage{bm}
\usepackage{amsmath}
\usepackage{amssymb}

\begin{document} 
 
\title{
Theoretical studies on off-axis phase diagrams and Knight shifts in UTe$_2$\/
---Tetra-critical point, d-vector rotation, and multiple phases--- }
\author{Kazushige Machida} 
\affiliation{Department of Physics, Ritsumeikan University, Kusatsu 525-8577, Japan} 

\date{\today}

\begin{abstract}
Inspired by recent remarkable sets of experiments on UTe$_2$: discoveries of the fourth horizontal internal transition line
running toward a tetra-critical point (TCP) at $H$=15T, the off-axis high field phases,
and abnormally large Knight shift (KS) drop below $T_{\rm c}$ for $H$$\parallel$$a$-magnetic easy axis, 
we advance further our theoretical work on the field ($H$)-temperature ($T$) 
phase diagram for $H$$\parallel$$b$-magnetic hard axis which contains a positive sloped $H_{\rm c2}$ departing from TCP.
A nonunitary spin-triplet pairing with three components explains these experimental facts simultaneously and consistently by assuming
that the underlying normal electron system with a narrow bandwidth characteristic to the Kondo temperature $\sim$60K 
unsurprisingly breaks the particle-hole symmetry. This causes a special invariant term in Ginzburg-Landau (GL) free energy functional
which couples directly with the 5f magnetic system, giving rise to the $T_{\rm c}$ splitting and ultimately to the positive sloped $H_{\rm c2}$ and
the horizontal internal transition line connected to TCP. The large KS drop can be understood
in terms of this GL invariance whose coefficient is negative and leads to a diamagnetic response where the 
Cooper pair spin is antiparallel to the applied field direction. The present scenario also accounts for the observed d-vector rotation phenomena and off-axis phase diagrams with
extremely high $H_{\rm c2}$$\gtrsim$70T found at angles in between the $b$ and $c$-axes and 
between the $bc$ plane and $a$-axis,
making UTe$_2$ a fertile playground for a topological superconductor.
\end{abstract}

\maketitle 
\section{Introduction}

Much attention has been focused on the recently found heavy Fermion superconductor UTe$_2$~\cite{ran,aoki0,review,review1}.
Since the upper critical fields $H_{\rm c2}$ for the principal directions all surpass the Pauli paramagnetic 
limiting field, it is regarded as a prime candidate for a spin-triplet superconductor that is quite rare in nature
except for UPt$_3$~\cite{taillefer,sauls,upt3,ohmi,yo,tsutsumi1}, UBe$_{13}$~\cite{ott,shimizu1,shimizu2}
and its Th-doped materials~\cite{Th1,shimizu3,Th} among abundant candidate materials.
In these superconductors, the phase diagrams of the field ($H$) and temperature ($T$) plane consists
of multiple distinctive phases. This is a hallmark of a triplet pairing because of its rich internal degrees of
freedom associated with the spin and orbital spaces in the pairing function.
This is exemplified by the textbook case of the p-wave superfluid $^3$He where the phase diagram 
in the $T$ vs $P$ (pressure) plane is divided into the ABM (A) and BW (B) phases~\cite{3he}.
Thus investigating the phase diagram in UTe$_2$ may be a fruitful route to identify the pairing symmetry realized.

UTe$_2$ has been discovered by Ran et al~\cite{ran} in 2019 with its transition temperature T$_{\rm c}=1.6$K.
Since then, there are a variety of experimental and theoretical works devoted to uncovering the
realized Cooper pair symmetry~\cite{review,review1}. The system belongs to the so-called heavy Fermion materials in general
with the enhanced electron effective mass and the Kondo coherent temperature $\sim$30K.
Initially it was thought that because of various similar physical properties it is a sister compound of
the ferromagnetic superconductors UGe$_2$, URhGe, and UCoGe, but no static long-ranged ferromagnetic 
order has been found. Instead, antiferromagnetic spin fluctuations are discovered~\cite{af1,af2,af3}.
As for the pairing symmetry~\cite{review,review1}, (A) the spin triplet pairing may be realized 
because of the Knight shift experiments showing the drop (unchanged)
of the spin susceptibilities for $H\parallel b$ and $c$-axes ($a$-axis). 
(B) The time reversal symmetry is broken.
(C) The chiral current on a sample edge is found. 
(D) The point nodes are oriented to the $a$-axis.

Since these experiments were mainly done by using T$_{\rm c}=1.6$K samples,
experiments using the new generation samples with T$_{\rm c}=2.1$K challenge some of these conclusions.
We are warned that some of the experimental facts are robust against sample quality, but some
of the conclusions may be changed. For example, the overall structure of the phase diagram in the $H$-$T$ plane 
for $H\parallel b$-axis with the field-reinforced positive sloped $H_{\rm c2}$ is hardly changed.
However, we need to further refine the important details of the existing phase diagram to
narrow down the possible pairing symmetry in UTe$_2$.

According to the recent report by Sakai et al~\cite{sakai} by using the new 
generation high-quality samples with $T_{\rm c}$=2.1K~\cite{sakai2.1K},
the three transition lines in the $H$-$T$ phase diagram ($H$$\parallel$$b$-axis) 
meet at a point of $H$=15T comprising a tetra-critical point (TCP).
This newly found fourth horizontal internal line runs almost parallel to the $T$-axis.
Together with the previous discovery of the double transition in the specific heat measurement~\cite{rosuel} 
corresponding to another two
transition lines, it is understood that at least three phases exist in the $H$-$T$ plane.
These experimental facts encourage us to further investigate this material with the high possibility of
a spin-triplet pairing state realized. In particular, we are urged to reproduce this characteristic $H$-$T$ phase diagram
with an appropriate pairing state, which could narrow down the possible pairing symmetry.

Here we put forth further our theoretical 
work~\cite{machida1,machida2,machida3,machida4,machida5} 
in the light of the above new experimental facts and reproduce the renewed phase diagrams 
with the horizontal fourth transition line for $H\parallel b$-axis~\cite{sakai} and extremely high $H_{\rm c2}$ for
off-axis directions~\cite{lewin}.
We also develop our thoughts based on the nonunitary triplet scenario to explain the revised
Knight shift (KS) experiments~\cite{matsumura} 
by using the high-quality samples with $T_{\rm c}$=2.1K together with the previous NMR results~\cite{ishida1,ishida2,ishida3,ishida4,ishida5,kinjo}.
They show that KS decreases below $T_{\rm c}$ for all three directions; $a$, $b$, and $c$-axes.
At first sight, this surprising result may be taken as a signature either for a spin singlet pairing
or for the B phase-like state in the superfluid $^3$He~\cite{3he}
where the d-vector has simultaneously the three components,
making KS decrease in all directions. In this paper, we argue that neither spin singlet scenario
nor the B phase-like state can consistently and coherently explain various compiled experimental
facts which are summarized as follows:\\
\noindent
(1) In the $H$-$T$ plane for $H$$\parallel$$b$-axis, the intermediate phase (MSC) is 
sandwiched between the low phase (LSC) and high field phase (HSC) at 15T~\cite{sakai,ishida5}.
The phase boundary of LSC-MSC comprises the fourth horizontal internal line
ending at TCP. 
And $H^b_{\rm c2}(T)$ has an unusual positive slope above 15T 
starting from TCP, namely $dH^b_{\rm c2}/dT>0$.\\
\noindent
(2) MSC is characterized by the flux flow state~\cite{sakai} in the $H$-$T$ plane for $H\parallel b$-axis.
The vortices in MSC are easily depinned under external currents, suggesting that an exotic vortex lattice is
formed in this field region around 15T, which coincides with the region where the d-vector is rotating mentioned below. \\
\noindent
(3) The highest upper critical field $H_{\rm c2}$ occurs at narrow angle regions centered at $\theta=35^{\circ}$
measured from the $b$-axis toward the $c$-axis~\cite{ran2,georg,helm,lonzarich}, 
and at $\phi\sim 10^{\circ}$ measured from the $bc$-plane toward the $a$-axis~\cite{lewin}, 
reaching a surprisingly large value of more than 60T
compared with $T_{\rm c}$=1.6K$\sim$2.1K.
This high field region appears just above the meta-magnetic transition $H_{\rm m}$.
This phenomenon is similar to, but distinctive from the so-called reentrant superconducting phase 
observed in the sister compound URhGe,
which appears in more wider angle region~\cite{urhge0,urhge1,urhge2}. 
For $H\parallel b$-axis $H^b_{\rm c2}(T)$ abruptly stops at $H_{\rm m}(\parallel b)=35$T.\\
\noindent
(4) In KS by the $^{125}$Te-NMR experiments, the decrease $\Delta K_a$ along the $a$-axis is extremely large~\cite{matsumura}
compared with the other two for the $b$-axis and $c$-axis, which are comparable, namely 
$|\Delta K_a|\gg |\Delta K_b|\sim|\Delta K_c|$~\cite{ishida1,ishida2,ishida3,ishida4,ishida5,kinjo}.
Since for the B phase~\cite{3he,mizushima} in $^3$He all three KS values are equal, corresponding to the
spin susceptibility $\chi_a=\chi_b=\chi_c=-{2\over 3}\chi_{\rm N}$ with $\chi_{\rm N}$ the Pauli spin
susceptibility in the normal state. The KS data are different from this fact and also from the spin-singlet state where 
$\chi_a=\chi_b=\chi_c=-\chi_{\rm N}$. \\
\noindent
(5) According to the NMR experiments~\cite{ishida2,ishida3} on the $T_{\rm c}$=1.6K samples, 
with increasing $H$$\parallel$$b$ $(\parallel$$c)$
the decrease of KS gradually diminishes, recovering to the normal $\chi_{\rm N}$ value.
This starts from 5T (a few T) and ends at 15T (5T)
above which the decrease of KS ceases completely. Namely, the d-vector gradually rotates from
 the parallel direction to the perpendicular direction relative to the external field. \\
\noindent
(6) Under pressure, the truly multiple phases~\cite{pressure1,pressure2,pressure3,pressure4,butch1,butch2,butch3} 
are observed and systematically evolving both as a 
function of $P$ and the field orientation. In particular, for $H\parallel a$-axis the
phase diagram consists of more than three phases, showing that the
order parameters must be three components.\\
\noindent
(7) There exist several important experiments to show the unconventional nature of the pairing symmetry~\cite{shimizu4}.
This includes the time-reversal symmetry breaking detected by polar Kerr effect~\cite{kaptulnik}, the broken chiral symmetry 
at sample edges seen by STM-STS spectroscopies~\cite{madhavan}, the gap structure with point nodes through various
bulk thermodynamic measurements~\cite{metz,kittaka}. They point to an unconventional pairing state realized in UTe$_2$.
Some of those data are under discussion~\cite {sonier1,shibauchi,matsuda} to finally pin down the precise pairing symmetry.

The main purposes of the present paper are threefold, the first
is to coherently explain these experimental facts with
minimum assumptions: the off-axis high field phases~\cite{ran2,georg,helm,lonzarich} with record high $H_{\rm c2}$,
in particular, the newly found phase diagrams~\cite{lewin} with the fields tilted from the $bc$-plane toward the $a$-axis,
which was not covered before~\cite{machida5}.
The second is to revisit the phase diagram for $H\parallel b$-axis containing
newly discovered intermediate phase around TCP at 15T 
with the horizontal fourth transition line~\cite{sakai} in order to sharpen our understanding.
The third is to explain the Knight shift experiment performed on the second generation
high-quality samples with $T_{\rm c}$=2.1K, which shows the unusual drop for $H\parallel a$-axis~\cite{matsumura}.
These should be understood consistently and coherently.  
This task was not done in our series of papers on this material~\cite{machida1,machida2,machida3,machida4,machida5}.

The arrangement of the present paper is as follows:
In the next section II, we outline our theoretical framework to analyze the
experimental data mentioned. Then, we revisit the $H$-$T$ phase diagram for $H\parallel b$-axis, emphasizing 
the horizontal fourth transition line toward TCP in Sec III. In Sec IV we reproduce the off-axis phase diagrams,
focusing on why the extremely high field $H_{\rm c2}$ is possible and why it is narrowly confined. 
In the following Sec V, we investigate the unusual Knight shift problems observed
in detail, which is one of the main highlights of the present paper.
We devote to the discussion in Sec VI and give a conclusion and summary in the final section.

The main difference from our series of papers~\cite{machida1,machida2,machida3,machida4,machida5} lies in the fact that
we further advance our theory, keeping the same theoretical framework.
The formulation is the same as before, however, we repeat it for completeness as a full paper which is self-contained.
In particular, we are now able to explain the newly found remarkable
experimental facts: the horizontal fourth internal transition line, the off-axis phase diagrams, and
the abnormally large Knight shift change for $H\parallel a$-axis simultaneously and coherently.
These new experiments become possible only by using newly synthesized high-quality samples
with $T_{\rm c}$=2.1K~\cite{sakai2.1K}.

\section{Ginzburg-Landau theory for three components}

\subsection{Assumptions}

We start with the most general Ginzburg-Landau (GL) theory for a spin triplet state~\cite{machida1,machida2,machida3,machida4,machida5} to analyze the
experimental data on UTe$_2$. We place three basic assumptions in the present paper:\\
\noindent
(1) We assume a nonunitary A-phase-like pairing state described by the complex $\bf d$-vector
${\bf d}(k)=\phi(k){\boldsymbol \eta}=\phi(k)({\boldsymbol \eta}'+i{\boldsymbol \eta}'')$ with
${\boldsymbol\eta}'$ and ${\boldsymbol \eta}''$ being real vectors.
$\phi(k)$ is the orbital part of the pairing function. \\
\noindent
(2) The pairing function is classified under the overall symmetry 
$SO(3)^{\rm spin}\times D_{2h}^{\rm orbital}\times U(1)^{\rm guage}$
with the spin, orbital, and gauge symmetry, respectively~\cite{machida0,annett},
assuming the weak spin-orbit coupling scheme~\cite{ozaki1,ozaki2}.
The justification for this weak spin-orbit coupling (SOC) scheme lies in the experimental fact that 
the d-vector rotation begins from the low fields, $\sim$1 T for the $c$-axis~\cite{ishida3}, and $\sim$5 T
and its gradual rotation is completed at 15T for the $b$-axis~\cite{ishida2}.
This indicates that the spin-orbit coupling, which locks the d-vector 
to crystalline lattices, is weak. We note that in the strong SOC scheme
the gradual d-vector rotation is not possible because the d-vector locking energy 
is infinitely strong. The d-vector may suddenly change when the two phases transform by a 
second-order or first-order phase transition
because the spin symmetry is reduced to the crystalline symmetry $D_{\rm 2h}$, 
yielding four one-dimensional irreducible representations~\cite{machida0,annett}.
These representations have distinct and 
non-degenerate transition temperatures
except for the accidental degeneracy.\\
\noindent
(3) To stabilize the nonunitary triplet pairing state~\cite{ramires}, we assume the
ferromagnetic fluctuations~\cite{ran,sonier,tokunaga1,tokunaga2,furukawa,tokunaga-prl} 
slower than the Cooper pair formation time. 
This allows us to introduce non-vanishing root-mean-square average $\sqrt{\langle M_a^2\rangle}$
along the magnetic easy $a$-axis. This effective symmetry breaker makes the
nonunitary state stable, otherwise unitary states obviously win over the nonunitary state. 
We do not need the static and/or spontaneous ferromagnetic transition,
which is absent in UTe$_2$ even in the samples with $T_{\rm c}$=2.1K~\cite{sonier1}.
According to the recent NMR experiment on high-quality samples,
Tokunaga et al~\cite{tokunaga-prl} discover extremely slow 
longitudinal magnetic fluctuations on their $T_2$ measurements
in the normal state although they do not specify whether or not 
 ferromagnetic or antiferromagnetic.
 
 We note that this postulation is supported by several previous examples similar to this situation:
 In UPt$_3$ the antiferromagnetic order at $T_{\rm N}$=7K lifts the degeneracy in the two-dimensional 
 E$_{\rm 1u}$ representation, but this antiferromagnetism  is not a  static and long-ranged order
 as checked by a variety of thermodynamic measurements, but only observed by the elastic neutron scattering~\cite{aeppli},
 meaning that the antiferromagnetic fluctuations are fast enough thermodynamically and slow enough 
 for neutron measurements. Yet they work as a symmetry breaker to split the SC transition temperature into two~\cite{sauls,upt3,ohmi,yo,tsutsumi1}.
 Another example comes from the electron-doped cuprates where the antiferromagnetic fluctuations are reported to 
yield the band folding brought by a magnetic wave vector observed by photoemission experiments~\cite{bandfolding}

The SO(3)$^{\rm spin}$ triple spin symmetry for the Cooper pair spin space
allows us to introduce a complex
three-component vectorial order parameter ${\boldsymbol \eta}=(\eta_a,\eta_b,\eta_c)$.
The spin space symmetry is weakly perturbed by the 5f localized moments of the U atoms through the ``effective'' 
spin-orbit coupling felt by the Cooper pairs in the many-body sense. 
Our framework based on the weak SOC is sufficiently flexible to include the strong SOC as a limit~\cite{anderson,blount,gorkov}.
As detailed in the following, the three components of the order parameter originally had the same transition temperatures
under the spin rotational symmetry, which is broken by either the applied field or the influence of the underlying magnetic subsystem.

\subsection{GL free energy functional}

Under D$_{2h}$$^{\rm orbital}$ symmetry, the GL free energy 
functional up to the quadratic order of $\boldsymbol \eta$ is expressed
by

\begin{eqnarray}
F=F^{(2)}_{\rm bulk}+F_{\rm grad} 
\label{free}
\end{eqnarray}

\noindent
where the free energy $F$ consists with the bulk part $F^{(2)}_{\rm bulk}$ and the gradient term $F_{\rm grad}$.
Each is given by

\begin{eqnarray}
F^{(2)}_{\rm bulk}=a_0(T-T_{\rm c0}){\boldsymbol \eta}\cdot{\boldsymbol \eta}^{\ast}+
b|{\bf M}\cdot{\boldsymbol \eta}|^2+i\kappa {\bf M}\cdot {\boldsymbol \eta}\times {\boldsymbol \eta}^{\ast}
\label{bulk}
\end{eqnarray}

\noindent
and 

\begin{eqnarray}
F_{\rm grad}=\sum_{\nu=a,b,c}\{K_a|D_x\eta_{\nu}|^2+K_b|D_y\eta_{\nu}|^2+K_c|D_z\eta_{\nu}|^2\}
\label{grad}
\end{eqnarray}

\noindent
where $b$ in Eq.~(\ref{bulk}) is a positive constant. 
The last invariant in Eq.~(\ref{bulk}) is essential for the following analyses
and results from the nonunitarity of the pairing function in the presence of the 
moment ${\bf M}$, which breaks the SO(3)$^{\rm spin}$ spin symmetry. 
$K_{a}$, $K_{b}$, and $K_c$ in Eq.~(\ref{grad}) are the effective masses along the $a$ $b$, and $c$-axes, respectively
under an applied field with the vector potential $\bf A$.
$D_i=-i\nabla_i+{2\pi\over \Phi_0}A_i$ is the gauge invariant derivative, with
$\Phi_0$ being the quantum flux and $A_i$ the vector potential component for the $i$-direction ($i=a, b, c$).

The magnetic coupling $\kappa$, which is a key parameter for  characterizing UTe$_2$,
 is originally estimated~\cite{mermin} as

\begin{eqnarray}
\kappa=T_{\rm c}{N'(0)\over N(0)}ln(1.14\Omega_{\rm c}/T_{\rm c}),
\label{kappa}
\end{eqnarray}

%$\kappa=T_{\rm c}{N'(0)\over N(0)}ln(1.14\omega_{\rm c}/T_{\rm c})$,

\noindent
where $N'(0)$ is the energy derivative of the normal density of states (DOS) $N(0)$, and 
$\Omega_{\rm c}$ is the energy cut-off.
This term results from the electron-hole asymmetry near the Fermi level. $\kappa$ indicates
the degree of this asymmetry.
This may be significant for a narrow band or the Kondo coherent band in the heavy Fermion
material UTe$_2$. We can estimate
$N'(0)/N(0)\sim 1/E_{\rm F}$ with the Fermi energy $E_{\rm F}$.
Because $T_{\rm c}$=2 mK and $E_{\rm F}$=1 K in $^3$He, $\kappa\sim10^{-3}$,
while for UTe$_2$ $T_{\rm c}\sim$1K and $E_{\rm F}\sim T_{\rm K}$ with the 
Kondo temperature $T_{\rm K}\sim$30 K~\cite{review},
$\kappa\sim10^{-1}$. 
We also note that the sign of $\kappa$ can be either positive or negative,
depending on the detailed energy dependence of DOS at the Fermi level because it is $\propto N'(0)$.
If $\kappa>0$ ($\kappa<0$), the $\uparrow$$\uparrow$ ($\downarrow$$\downarrow$) pair appears at higher $T$.
Thus, the KS remains unchanged (decreases) below $T_{\rm c}$.
We will find later that $\kappa<0$ in UTe$_2$. That is, we now determine the sign of $\kappa$

%$\eta_{\pm}={1\over \sqrt2}(\eta_b\pm i\eta_c)$

%\begin{eqnarray}
%\eta_{\pm}={1\over \sqrt2}(\eta_b\pm i\eta_c)
%\label{eta}
%\end{eqnarray}

We introduce $\eta_{\pm}=(\eta_b\pm i\eta_c)/\sqrt2$
for ${\bf M}=(M_a,0,0)$ where we define the $a$-axis as the magnetic easy axis. 
$\eta_{+}$ ($\eta_{-}$) corresponds to the spin $\uparrow$$\uparrow$ ($\downarrow$$\downarrow$) 
 pair or the A$_1$(A$_2$) phase.
The spin quantization axis is defined relative to the ${\bf M}$ direction, i.e., the
magnetic easy $a$-axis. Because of the magnetic coupling term 
$i\kappa \bf{M}\cdot \boldsymbol \eta\times\boldsymbol \eta^{\ast}$, the spin direction for the
Cooper pair may change. Here it is convenient to introduce the Cooper pair spin moment direction
$\boldsymbol S$ as

\begin{eqnarray}
\boldsymbol S=i {\boldsymbol \eta\times\boldsymbol \eta^{\ast}\over |\boldsymbol \eta|^2}.
\label{spin}
\end{eqnarray}

\noindent
We notice the case: when the direction $\boldsymbol S$ is fixed, the magnetization  ${\bf M}$ direction is chosen
to lower the magnetic coupling energy $i\kappa {\bf M}\cdot {\boldsymbol \eta}\times {\boldsymbol \eta}^{\ast}$.
From Eq.~(\ref{bulk}) the quadratic term $F^{(2)}_{\rm bulk}$ becomes
%$F^{(2)}=\alpha_0(T-T_{\rm c \uparrow})|\eta_{+}|^2+\alpha_0(T-T_{\rm c \downarrow})|\eta_{-}|^2$
%$F^{(2)}=\alpha_0\{(T-T_{\rm c1})|\eta_{+}|^2+(T-T_{\rm c2})|\eta_{-}|^2+(T-T_{\rm c3})|\eta_{a}|^2\}$

\begin{eqnarray}
F^{(2)}_{\rm bulk}=a_0\{(T-T_{\rm c1})|\eta_{+}|^2+(T-T_{\rm c2})|\eta_{-}|^2\nonumber \\
+(T-T_{\rm c3})|\eta_{a}|^2\}
\label{bulk2}
\end{eqnarray}

\noindent
with 

\begin{eqnarray}
T_{\rm c 1,2}(M_a)=T_{\rm c0} \pm{\kappa\over a_0}M_a,\nonumber\\
T_{\rm c 3}(M_a)=T_{\rm c0} -{b\over a_0}M^2_a.
\label{tc1}
\end{eqnarray}

\noindent
Although the actual second transition temperatures are modified 
by the fourth order GL terms~\cite{machida1,machida2,machida3},
we ignore this correction and maintain the expressions for the
transition temperatures for clarity.

The root-mean-square average  $\sqrt{\langle M_a^2\rangle}$ of 
the FM fluctuations along the magnetic easy $a$-axis
was simply denoted by $M_a$, shifting the transition temperature $T_{\rm c0}$ and split it into $T_{c1}$,
$T_{c2}$, and $T_{c3}$ given above.
According to this, $T_{c1}$ ($T_{c2}$) increases (decreases) linearly as a function of $M_a$,
whereas $T_{c3}$ decreases quadratically as $M^2_a$ from the degeneracy point $M_a=0$.
The three transition lines meet at $M_a$=0, where the 
three components $\eta_i$ ($i=+,-,a$) are all degenerate. Thus, away from the degenerate point 
at $M_a$=0, the A$_0$ phase beginning at $T_{\rm c3}$ quickly disappears from the phase diagram in general.
Below $T_{\rm c2}$, the two components $\eta_{+}$ and $\eta_{-}$ coexist, symbolically denoted by
A$_1$+A$_2$. Note that, because  their transition temperatures are different,
A$_1$+A$_2$ is not the so-called A-phase, which is unitary, but here is generically nonunitary
except at the degenerate point $M_a$=0 where the totally symmetric phase is achieved with the 
time reversal symmetry preserved. This occurs under pressure at the critical pressure 
$P_{\rm cr}$=0.15GPa for $T_{\rm c}$=2.1K samples when $T_{\rm c1}$=$T_{\rm c2}$.
 The A$_1$+A$_2$ phase is the so-called distorted A phase~\cite{3he}. 
Similarly, below $T_{\rm c3}$, all the components coexist; A$_1$+A$_2$+A$_0$ is realized.
The naming such as A$_1$,  A$_2$, and  A$_0$ is retained even when the spin quantization 
axis changes according to the d-vector rotation under fields.
Thus  A$_1$ and A$_2$ are meant to distinguish between
the spin $\uparrow$$\uparrow$ pair and spin $\downarrow$$\downarrow$
pair with respect to the spin quantization axis.
%In the following discussions, we consider a case in which the two components $\eta_{+}$ and $\eta_{-}$
%are non-vanishing, ignoring the third component $\eta_{a}$ because, 
%under ambient pressure, UTe$_2$ exhibits the two phases LSC and HSC, corresponding to 
% $\eta_{+}$ and $\eta_{-}$, respectively.
% The current UTe$_2$ samples with $T_{\rm c}$=1.6 K exhibit
% a single transition under $H$=0 because $T_{\rm c2}<0$, as evidenced by the significant
% residual density of states. The second transition is
 %only realized at a finite field for $H\parallel$ b-axis.
%However, we expect that the second transition $T_{\rm c2}>0$ might be realized for
 %samples with $T_{\rm c}$=2 K without the residual density of states.
 %Note that, under pressure, the third component becomes relevant~\cite{machida2}
% (see also the Appendix for the third transition).
Hereinafter, we redefine the notation $\kappa/a_0\rightarrow \kappa$.
 
\subsection{Upper critical field}

To construct the phase diagrams for various field orientations,
we need to derive the $H_{\rm c2}$ expression.
This can be done by starting with the free energy functional in Eq.~(\ref{free})
where Eq.~(\ref{bulk}) is rewritten as in Eq.~(\ref{bulk2}).
The form of Eq.~(\ref{grad}) shows
that the $H_{\rm c2}$ for the three components each starting at $T_{\rm c j}$ ($j=1,2, 3$)
given by Eq.(\ref{tc1}) intersect
each other, never avoiding or leading to a level repulsion. The level repulsion may occur 
for the pairing states belonging to multi-dimensional representations
 (see for example [\onlinecite{repulsion1,repulsion2,repulsion3,repulsion4}] in UPt$_3$).
Thus, each component is independent within the quadratic terms.
The total GL free energy density $F$, consisting with  Eqs.~(\ref{grad}) and (\ref{bulk2}) 
under the external magnetic field $H$
in terms of the superconducting order parameter $\eta_{\pm}$ is given by

\begin{eqnarray}
F&=&\sum_{i=\pm}\{a_0(T-T_{{\rm c},i})|\eta_{i}|^2  \nonumber \\ 
& &+K_{a}|D_x\eta_i|^2+K_{b}|D_y\eta_i|^2+K_c|D_z\eta_i|^2\}.
\label{GL}
\end{eqnarray}

\noindent
The variation with respect of $\eta_i^{\ast}$ results in  

\begin{eqnarray}
a_0(T-T_{\rm c})\eta_i+(K_{a}D_x^2+K_{b}D_y^2+K_cD_z^2)\eta_i=0.
\label{osci}
\end{eqnarray}

\noindent
The upper critical field $H_{\rm c2}$ is given as the lowest eigenvalue of the linearized 
GL equation or Schr\"odinger type equation of a harmonic oscillator~\cite{tinkham} as,

%\begin{eqnarray}
%H^{(+)}_{{\rm c2},j}(T)=\alpha^j_0\{T_{\rm c1}(M)-T\}\nonumber \\ 
%H^{(-)}_{{\rm c2},j}(T)=\alpha^j_0\{T_{\rm c2}(M)-T\}
%\label{hc2}
%\end{eqnarray}

\begin{eqnarray}
H^{(\pm)}_{{\rm c2},j}(T)=\alpha^j_0(T_{\rm c0} \pm\kappa M_a-T)
\label{hc2}
\end{eqnarray}

\noindent
with $j$=$a$, $b$, and $c$-axis. We have introduced, 
$\alpha^{a}_0={\Phi_0\over 2\pi \sqrt{K_bK_{c}}}a_0$,
$\alpha^{b}_0={\Phi_0\over 2\pi \sqrt{K_cK_{a}}}a_0$,
and $\alpha^{c}_0={\Phi_0\over 2\pi \sqrt{K_aK_{b}}}a_0$.

%\begin{eqnarray}
%\alpha^{a}_0&=&{\Phi_0\over 2\pi \sqrt{K_bK_{c}}}a_0,\qquad
%\alpha^{b}_0={\Phi_0\over 2\pi \sqrt{K_cK_{a}}}a_0,\nonumber \\ 
%\alpha^{c}_0&=&{\Phi_0\over 2\pi \sqrt{K_aK_{b}}}a_0.
%\label{mass}
%\end{eqnarray}

\noindent
These coefficients determine the initial slopes of the upper critical fields.
$H^{(+)}_{{\rm c2},j}$ and $H^{(-)}_{{\rm c2},j}$ are the upper critical fields for 
 $\uparrow$$\uparrow$ and $\downarrow$$\downarrow$ pairs, or the A$_1$ and A$_2$ phases.

Expressing Eq. (\ref{hc2}) in a form by suppressing the index $j$,
we obtain in a general form:

\begin{eqnarray}
H_{{\rm c2}}-\alpha_0\kappa M(H_{\rm c2})=\alpha_0(T_{\rm c0}-T).
\label{gl}
\end{eqnarray}

\noindent
The right-hand side of Eq. (\ref{gl}) is now

\begin{eqnarray}
H_{\rm c2}^{\rm orb}(T)=\alpha_0(T_{\rm c0}-T)
\label{heff}
\end{eqnarray}

\noindent
for an unperturbed upper critical field owing to the orbital depairing limit
with $T_{\rm c0}$ whose maximum value is given by $H_{\rm c2}^{\rm orb}(T=0)=\alpha_0T_{\rm c0}$.
On the left-hand side of Eq.(\ref{gl}) we define the effective field $H_{\rm eff}$ by

\begin{eqnarray}
H_{\rm eff}=H_{{\rm ext}}-\alpha_0\kappa M(H_{\rm ext}).
\label{heff}
\end{eqnarray}

\noindent
This implies that the external field $H_{{\rm ext}}$ is reduced by 
$\alpha_0\kappa M(H_{\rm ext})$, a situation similar to that in CeRh$_2$As$_2$~\cite{machida4}
and also somewhat to the Jaccarino and Peter mechanism~\cite{jp} in a spin-singlet superconductor. 
The upper bound of the orbital depairing field of $H_{\rm c2}^{\rm orb}(T)$
for the $a$-axis, for example, is determined by 

\begin{eqnarray}
H_{\rm c2}^{\rm orb}(T\rightarrow0)=
\alpha^a_0T_{\rm c0}={\Phi_0\over 2\pi \sqrt{K_bK_{c}}}a_0T_{\rm c0}.
\label{h0}
\end{eqnarray}

\noindent
This is given by the expression  in the clean limit: $H_{\rm c2}^{\rm orb}(T)={\Phi_0/ 2\pi \xi^2}$, 
where the coherence length $\xi={\hbar v_{\rm F}/ \pi T_{\rm c0}}$.
At $H_{\rm c2}^{\rm orb}(T=0)$, the inter-vortex distance becomes comparable to
the core size $\xi$. From now on, we suppress the subscript ext;
thus, $H_{\rm ext}\rightarrow H$.

At $T=0$, the absolute value of $H_{\rm eff}$ is bounded by

\begin{eqnarray}
|H_{\rm c2}-\alpha_0\kappa M(H_{\rm c2})|\leq H^{\rm orb}_{\rm c2}(T=0)=\alpha_0T_{\rm c0}
\label{bound}
\end{eqnarray}

\noindent
for $H_{\rm c2}(0)$ to be a solution of Eq.~(\ref{gl}). The right-hand side is determined
by the material parameters in terms of the Fermi velocity $v_{\rm F} $ through the coherent length $\xi$ and the
transition temperature $T_{\rm c0}$.
Thus,  the absolute upper limit (AUL) $H^{\rm AUL}_{\rm c2}(0)$ can be enhanced at $T\rightarrow 0$
over $H^{\rm orb}_{\rm c2}(T=0)$, namely,

\begin{eqnarray}
H^{\rm AUL}_{\rm c2}(T)\ge H^{\rm orb}_{\rm c2}(T).
\label{aul}
\end{eqnarray}

\noindent
$H^{\rm AUL}_{\rm c2}(T)$ is a clue to understand the extremely high upper critical
field in UTe$_2$ as seen shortly.

\section{Phase diagram for $H\parallel b-$axis with tetra-critical point}

\subsection{Construction of the phase diagram}

\begin{figure}
\begin{center}
\includegraphics[width=6cm]{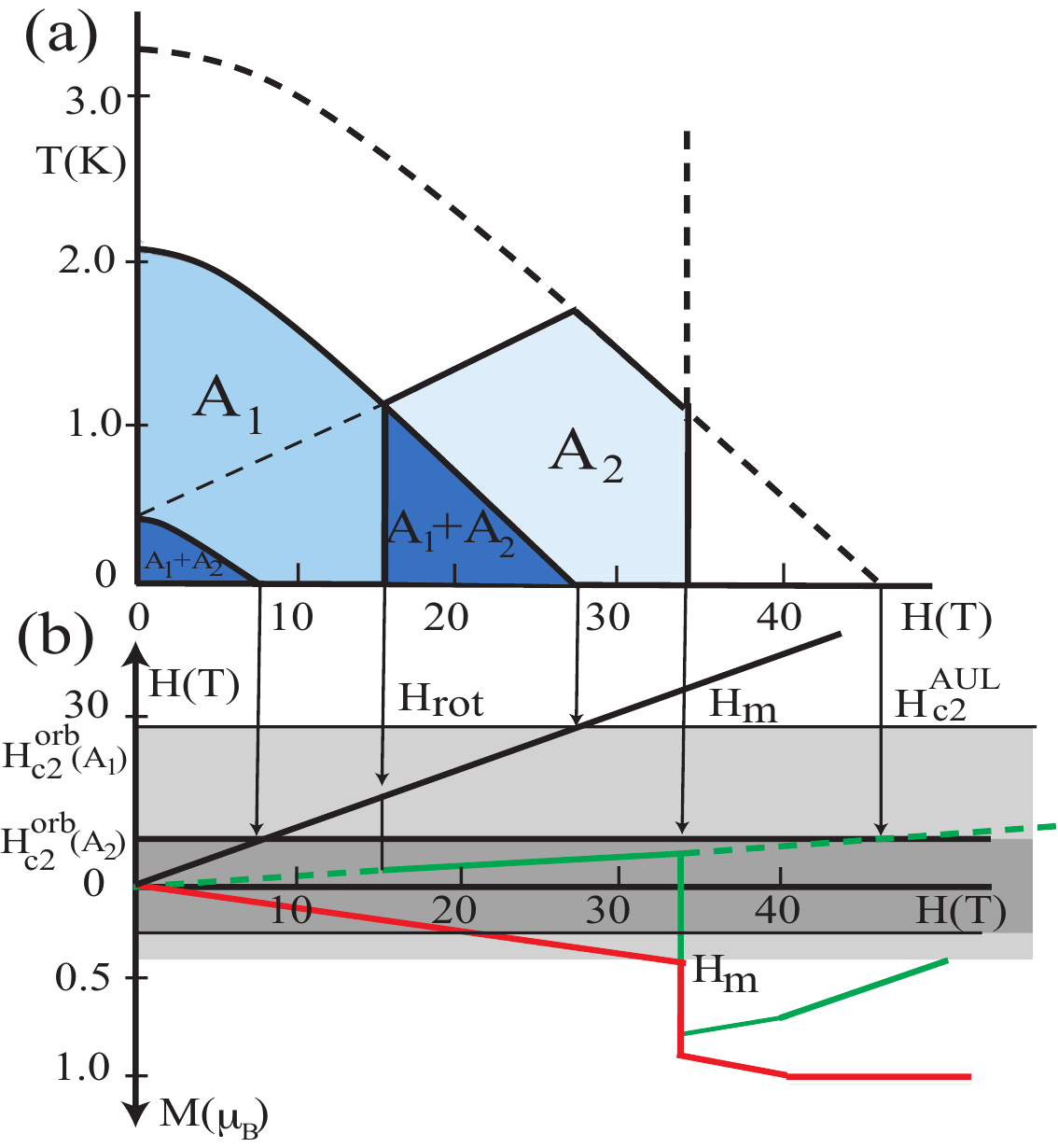}
\end{center}
\caption{\label{fig1}
Upper panel (a): Constructed phase diagram for $H\parallel b$-axis in the $T$ vs $H$ plane.
The A$_1$, the mixture A$_1$+A$_2$, and A$_2$ correspond to LSC, MSC, and HSC respectively
along the $H$ axis. Note that the A$_2$ exists in the lowest $H$ and lowest $T$ around the origin.
The dotted lines are not realized. Lower panel (b): $H_{\rm eff}(H)$ is indicated by the green curve, which is derived 
from the observed magnetization curve $M_b(H)$ denoted by the red line. $H_{\rm rot}$ is the field where
the d-vector rotation completes, starting around 5T. The two horizontal bands colored by light and dark gray
indicate the allowed field regions  $H^{\rm orb}_{\rm c2}({\rm A}_1)$  and $H^{\rm orb}_{\rm c2}({\rm A}_2)$
 for the A$_1$ and A$_2$ phases respectively. The dotted green line is not realized.
 The green line reenters the allowed region at $H^{\rm AUL}_{\rm c2}({\rm lower})$=58T and exits at $H^{\rm AUL}_{\rm c2}({\rm upper})$=73T in the higher field (see Fig.~\ref{fig5}).
}
\end{figure}

We start to investigate the phase diagram for $H\parallel b$-axis, which has the
richest structure with a tetra-critical point among three principal directions. To understand it, let us first examine the
magnetization curve $M(H\parallel b)=M_{b}(H)$ measured by Miyake et al~\cite{miyake}.  
As shown in the lower panel of Fig.~\ref{fig1} by the red line,
$M_{b}(H)$ exhibits a meta-magnetic first-order phase transition at $H_{\rm m}=34$T with the large magnetization jump
$\sim 0.5\mu_{\rm B}$/U-atom. Correspondingly, as defined in Eq.~(\ref{heff}) the effective magnetic field $H_{\rm eff}$ exerted to the
conduction electrons, which is denoted by the green curve is reduced strongly
compared with the external field $H_{\rm ext}$ shown there.
We have fixed the parameter values as follows:
$\alpha_0=27$T/K taken from the experimental data determined by 2.1K samples~\cite{initial}.
$\kappa=2.7$K/$\mu_{\rm B}$. This yields $T_{\rm c1}$=2.1K, $T_{\rm c0}$=1.25K, and $T_{\rm c2}$=0.44K
under the assumption $M_a=0.3\mu_{\rm B}$/U-atom.

For the A$_1$ phase with the transition temperature $T_{\rm c1}$=2.1K,
the allowed region for their $H^{\rm orb}_{\rm c2}(A_1)=\alpha_0 T_{\rm c1}$ in Eq.~(\ref{h0}) 
denoted by the light gray band is 
exceeded at $\sim$22T. Thus the  A$_1$ phase ceases to exist beyond this field at $T=0$.

On the other hand,  the A$_2$ phase with $T_{\rm c2}$=0.4K which has a narrower 
allowed region (dark gray band)  defined by $H^{\rm orb}_{\rm c2}(A_2)=\alpha_0 T_{\rm c2}$ 
compared with that for the A$_1$ phase by a factor $T_{\rm c2}$/$T_{\rm c1}$=0.2.
Thus this A$_2$ phase terminates at a lower field around 7T. However, 
once the d-vector rotation occurs at $H_{\rm rot}$=15T 
so that the Cooper pair spin ${\boldsymbol S}$ becomes parallel to the
$b$-axis to reduce the Zeeman energy, then it appears again 
and grows with $T_{\rm{c2}}=T_{\rm{c0}}+\kappa M_b(H)$. Note the switching $-M_a(H)$ to $+M_b(H)$
in Eq.~(\ref{tc1}) upon the d-vector rotation.  
As displayed in the upper panel of Fig.~\ref{fig1}, the intersection point at $H_{\rm tetra}$=15T where
the four second order phase transition lines meet to form a tetra-critical point (TCP).
The rising $T_{\rm{c2}}(M_b)$ line terminates at the meeting point with the
curve starting at $H_{\rm c2}^{\rm AUL}$. This defines the absolute upper limit of  
$H_{\rm c2}^{\rm orb}$(A$_2$) for the A$_2$ phase beyond which the green curve is outside the
allowed region colored by the dark gray.

Note that $H_{\rm c2}^{\rm AUL}(T)$ for $H_{\rm m}<H_{\rm c2}^{\rm AUL}(T)< H_{\rm c2}^{\rm AUL}(T=0)$ 
does not exist because the green curve is outside the allowed region indicated by the dark gray above $H_{\rm m}$.
Therefore, it exists only in the field region between the intersection point and the point below $H_{\rm m}$.
We also notice that the green line $H_{\rm eff}$ returns eventually to the allowed region in higher fields
denoted by $H_{\rm c2}^{\rm AUL}({\rm upper})$ and $H_{\rm c2}^{\rm AUL}({\rm lower})$,
but in this field the DOS available for superconductivity is below the normal DOS, thus is not eligible for the
A$_2$ phase to reappear as explained shortly.

\subsection{Tetra-critical point with horizontal fourth line}

\begin{figure}
\begin{center}
\includegraphics[width=6cm]{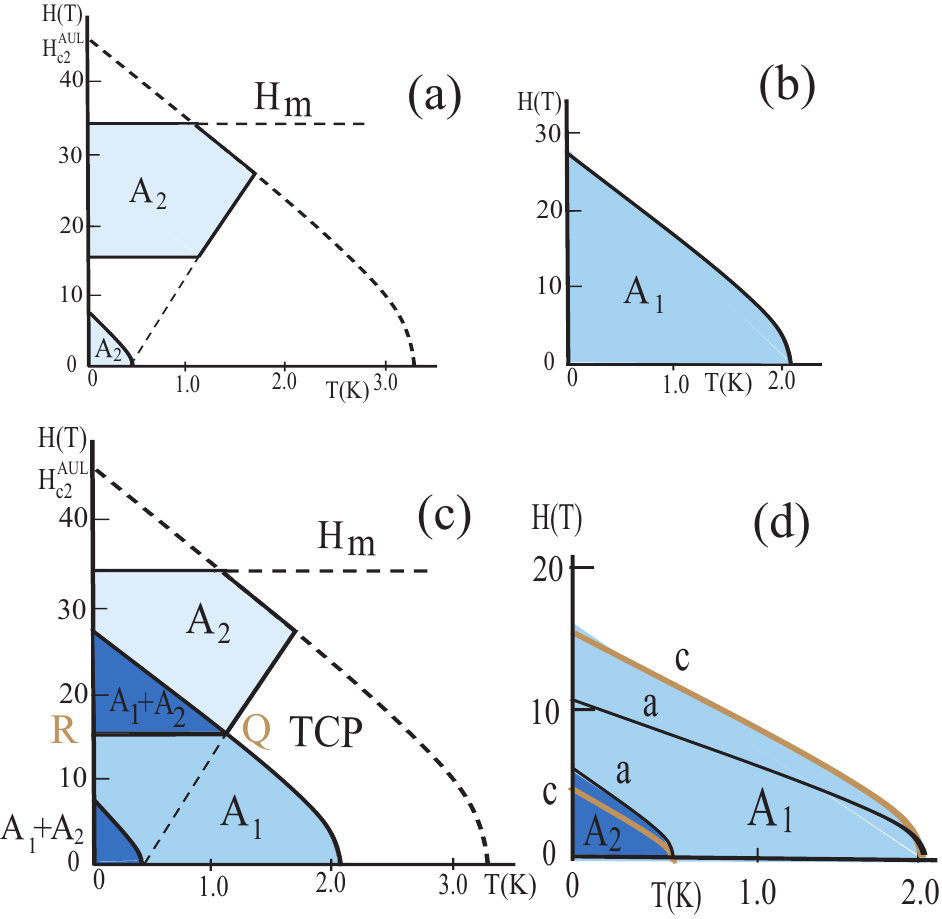}
\end{center}
\caption{\label{fig2}
(a) and (b) indicate the $H$-$T$ phase diagrams for the A$_1$ and A$_2$
phases separately. From (a), it is understood that the A$_2$ phase is reentrant. The isolated high 
field part conforms to HSC.
(c) shows the combined total phase diagram for $H\parallel b$-axis in the $H$-$T$ plane.
 The tetra-critical point (TCP) at $H_{\rm TCP}$=15T
consists of the horizontal forth line above which the A$_2$ phase reappears, the positive
sloped $H_{\rm c2}$ belongs to A$_2$, and the negative sloped internal line comes from A$_1$.
The dotted lines are not realized, displayed for guidance. The expected straight dotted line starting at $T_{\rm c2}$=0.4K
is not realized. (d) shows the phase diagrams for $H\parallel a$-axis and $H\parallel c$-axis, containing the small pocket
region for the A$_2$ phase.
}
\end{figure}

As seen from Fig.~\ref{fig2}(c),
along $H_{\rm c2}$ starting at $T_{\rm c1}$=2K, the A$_1$ phase characterized by 
$\boldsymbol \eta_b+i\boldsymbol\eta_c$ becomes non-vanishing.
At the tetra-critical point (TCP), $\boldsymbol \eta_a+i\boldsymbol \eta_c$ emerges associated with $T_{\rm c2}$.
Thus along the $H_{\rm c2}$ line just at TCP where $\boldsymbol \eta_a$=0, the d-vector can complete the rotation
from $\boldsymbol \eta_a+i\boldsymbol \eta_c$ to $\boldsymbol \eta_b+i\boldsymbol \eta_c$.
Simultaneously, above this particular field $H_{\rm TCP}$=15T in $T=0$,
this A$_2$ phase enters the allowed region denoted by the dark gray as seen 
from the green line of the lower panel of Fig.~\ref{fig1}.
Thus the fourth internal transition line toward TCP is always horizontal in the
$H$-$T$ phase diagram. 
It is true for the phase diagrams for $H\parallel b$-axis under pressure
(see Figs.~\ref{fig9}(b1), (b2), and (b3)).

Here it is important to notice that if the crossing of the second order phase transition lines is due to the
degeneracy of the orbital part of the pairing function $\boldsymbol \eta_i$ with $i$ being the orbital index,
there always exists the so-called gradient coupling term 
$(D_i \eta_j)^{*}(D_j \eta_i)$ with $D_i$ the gauge invariant derivative in the GL functional~\cite{repulsion1,repulsion2,repulsion3,repulsion4}.
This invariant inevitably leads to ``the level repulsion'' so that the four transition lines are 
always anti-crossing. This means that since the experimental data exhibit no anti-crossing,
the degeneracy comes from the spin part of the pairing function. 
We emphasize that this experimental observation of the tetra-critical point is crucial in choosing the
proper pairing function.

Therefore, as shown in Fig.~\ref{fig2} (b)
the A$_1$ phase is simply overlapped by  
the A$_2$ phase in Fig.~\ref{fig2} (a) to complete the overall phase diagram
shown in Fig.~\ref{fig2} (c). 
This means physically that the two A$_1$ and A$_2$ phases
interact weakly, allowing us to treat it in the weak coupling GL framework
as we apply. The forth order term of the interaction between them in our GL functional 
can be neglected.
The overlapping part between $H$=15T and 27T 
denoted by the darker color in Fig~\ref{fig2}(c) consists
of the mixture of the A$_1$ phase and A$_2$ phases.

\subsection{Horizontal fourth line}

According to the experimental data by Sakai, et al~\cite{sakai},
the fourth phase transition line toward TCP is almost horizontal.
Here we elaborate its reasons why it is so:

\noindent
(1) We first notice that this transition line comes from  the order parameter $\Delta_{b\downarrow}$
with the down spin of the Cooper pair whose quantization axis $\parallel$$b$-axis.
This state only senses the $M_b$ component, which is a fundamental assumption
throughout the present paper, even during the d-vector rotation process.

\noindent
(2) Thus the green curve for $M_b(H)$ in Fig.~\ref{fig2} is relevant after the d-vector rotation is completed.

\noindent
(3) The equi-$M_b$ lines  are all horizontal at the lower temperature region in Fig.~\ref{fig2}(c),
meaning that Fig.~\ref{fig1}(b) is basically applicable for finite temperatures.

\noindent
(4) The position R in Fig.~\ref{fig2}(c) at $T$=0 for $H_{\rm rot}$ with
$M_b=M_b(H_{\rm rot})$ must connect to the point Q with the same 
$M_b$ which is on the $T_{\rm c2}$ line with $T_{\rm c2}=T_{\rm c0}+\kappa M_b(H_{\rm rot})$.
 
\noindent
(5) Namely, Q satisfies the condition for the phase boundary.
By increasing $H$ under the fixed $T$, Q  becomes the onset for the A$_2$ phase.

\noindent
(6) Moreover the Q point is precisely the point which allows the completion of the
d-vector rotation because  the three components of the order parameter are
all degenerate there.

In Appendix A, we examine the phase diagram for the $T_{\rm c}$=1.6K samples.

%\subsection{The phase diagram tilted by small angles away from the $b$-axis}

\section{Off-axis high field phases}
%see Note 23.8.9 for $H_{\rm c2}^{\rm AUL}(\theta)$ expression.

When the field is tilted from the $b$-axis by a few degrees either toward the $a$-axis
or the $c$-axis, the above phase diagram in Figs.~\ref{fig1} and \ref{fig2}(c) is smoothly modified.
This is because the projection of $M_b(H)$ onto the external field is reduced,
thus correspondingly $H_{\rm eff}$ becomes smaller, resulting in $H^{\rm AUL}_{\rm c2}<H_{\rm m}$.
The whole part of the $A_2$ phase appears below $H_{\rm m}$
while a part of the $A_2$ phase is masked by $H_{\rm m}$ for $H\parallel b$-axis in Fig~\ref{fig2}.

\subsection{$H$-$T$ phase diagram in $b\rightarrow c$}

When the field is tilted by the larger angles $\theta$ measured from the $b$-axis toward the $c$-axis, the high field
phase, or HSC appears above $H_{\rm m}$ for a certain angle region. 
The forgoing arguments for $H\parallel b$-axis are extended to this case by changing

\noindent
(1) the magnetization curve $M_b(H)$ taking the projection onto the applied field
(see Fig. 8(b) in Ref.~\onlinecite{machida3}).

\noindent
(2) the allowed region $\pm \alpha_0(\theta)T_{\rm c2}$ where 

%$$\alpha_0(\theta)=\alpha_{0b}\cos\theta+\alpha_{0c}(1-\cos\theta)$$

\begin{eqnarray}
\alpha_0(\theta)=\alpha^b_{0}\cos\theta+\alpha^c_{0}(1-\cos\theta)
\label{tc}
\end{eqnarray}

\noindent
with $\alpha^b_{0}$ ($\alpha^c_{0}$) the value for the $b$ ($c$)-axis
by taking into account the observed $H_{\rm c2}$ anisotropy for the LSC phase due to the Fermi velocity change.
$\alpha^b_{0}$ is estimated by the initial slope of $H_{\rm c2}$ at $T_{\rm c}$=2.1K
as $\alpha^b_{0}=-dH_{\rm c2}/dT=-27$T/K ($\alpha^c_{0}=-6$T/K), which is known experimentally\cite{initial}.
We assume the angle dependence of   $\alpha(\theta)$ to be true for the high field $A_2$ phase too
because the Fermi velocity is the same for both phases.
Formally, $H^{\rm AUL}_{\rm c2}(\theta)$ is given by

%$$H^{\rm AUL}_{\rm c2}(\theta)={\alpha_0(\theta)T_{\rm c0}\over {1-\alpha_0(\theta)\kappa\chi_b\cos\theta}}$$
%$$={\alpha_0(\theta)T_{\rm c0}\over {1-\kappa\chi_b\{\alpha_{0b}\cos\theta+\alpha_{0c}(1-\cos\theta)\}\cos\theta}}$$
%$$={\alpha_0(\theta)T_{\rm c0}\over {1-\kappa\chi_b\{\alpha_{0b}\cos^2\theta+\alpha_{0c}\sin^2\theta-\alpha_{0c}(1-\cos\theta)\}}}$$

\begin{eqnarray}
H^{\rm AUL}_{\rm c2}(\theta)={\alpha_0(\theta)T_{\rm c0}\over {1-\alpha_0(\theta)\kappa\chi_b\cos\theta}} \label{tc}
\end{eqnarray}

%\begin{eqnarray}
%H^{\rm AUL}_{\rm c2}(\theta)={\alpha_0(\theta)T_{\rm c0}\over {1-\alpha_0(\theta)\kappa\chi_b\cos\theta}} \nonumber \\
%={\{\alpha_{0b}\cos\theta+\alpha_{0c}(1-\cos\theta)\}T_{\rm c0}\over {1-\kappa\chi_b\{\alpha_{0b}\cos\theta+\alpha_{0c}(1-\cos\theta)\}%\cos\theta}}
%\label{tc}
%\end{eqnarray}

\noindent
where $\chi_b$ is the susceptibility along the $b$-axis.
This means that the extra-enhancement of $H^{\rm AUL}_{\rm c2}(\theta)$
may come from this renormalization in the denominator.
However in the following we maintain to use the simplest formula 
without the renormalization for our analysis.

\noindent
(3) the d-vector rotation field is assumed to vary as 

\begin{eqnarray}
H_{\rm rot}(\theta)={H_{\rm rot}(\theta=0)\over \cos\theta},
\label{hrot}
\end{eqnarray}

\noindent
reflecting the fact that the projection of $M_b$ onto the external filed matters
since the coupling energy is proportional to $\kappa {\bf M}_b(H)\cdot {\boldsymbol \eta}\times {\boldsymbol \eta}^{\ast}$.
This idea is the same as $H_{\rm m}(\theta)=H_{\rm m}(\theta=0)/\cos(\theta)$ for the metamagnetic transition,
which is known to hold experimentally~\cite{review}.

\begin{figure}
\begin{center}
\includegraphics[width=5cm]{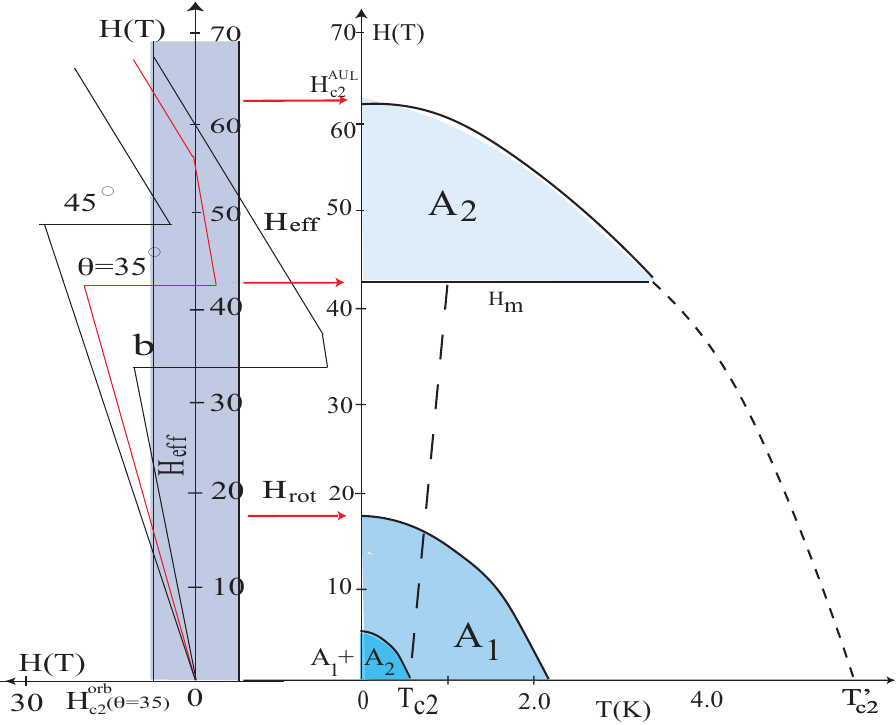}
\end{center}
\caption{\label{fig3}
The constructed phase diagram in the $H$-$T$ plane for $\theta=35^{\circ}$.
The left side panel: The allowed region denoted by the gray band with  $\pm H^{\rm orb}(\theta=35^{\circ})$=18T.
$H_{\rm eff}(\theta=35^{\circ})$ is showed by the red line together with those with 
$\theta=0^{\circ}$($\parallel b$) and $40^{\circ}$ for reference.
Above $H_{\rm m}$, it enters the allowed region up to $H^{\rm AUL}_{\rm c2}$, 
stabilizing the A$_2$ phase as shown in the
right panel where the whole $H$-$T$ phase diagram is displayed including the A$_1$ phase.
$H_{\rm rot}(\theta=35^{\circ})=16{\rm T}/\cos35^{\circ}=19.5$T just fails to yield
the tetra-critical point. There is no MSC at this angle.}
\end{figure}

We display an example of this procedure  for $\theta=35^{\circ}$ in Fig.~\ref{fig3}
where the panel on the left hand side shows the effective field $H_{\rm eff}$ as a function of the
external field $H$ by the red curve
relative to the allowed region and the resulting phase diagram is in the right panel.
Since the red curve for the $A_2$ phase comes back in the allowed region above $H_{\rm m}$ 
and stays up to $H_{\rm c2}^{\rm AUL}$, the $A_2$ phase reentrants above $H_{\rm m}$
while in the lower field, it disappears.
Here the meta-magnetic transition is important for the $A_2$ phase  to reappear
because $H_{\rm eff}$ is greatly reduced there.
Also the extra DOS $\gamma(H)(>\gamma_{\rm N})$ becomes available just above $H_{\rm m}$
as explained later.

\begin{figure}
\begin{center}
\includegraphics[width=5cm]{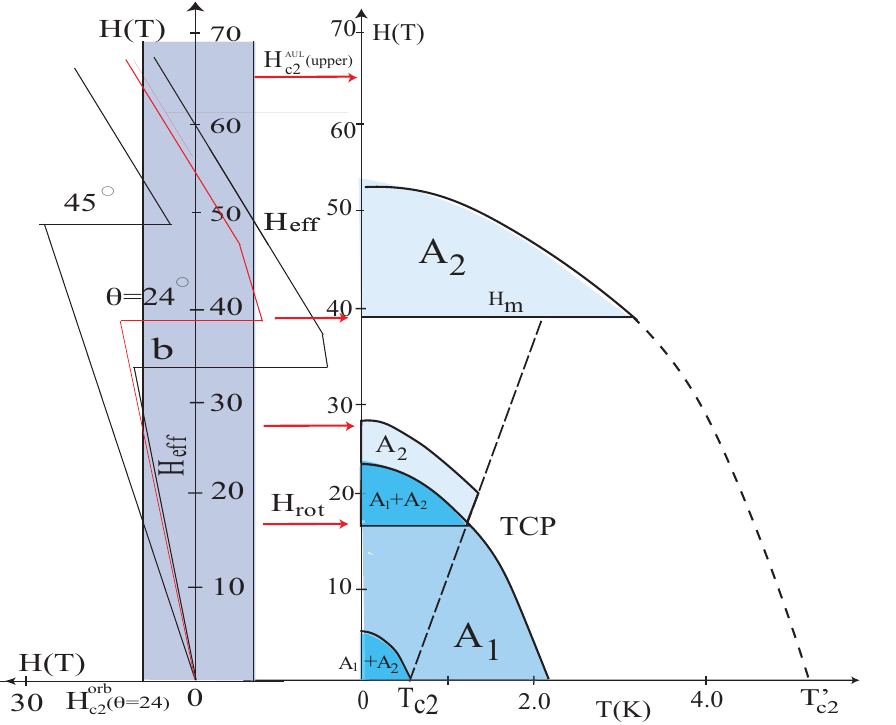}
\end{center}
\caption{\label{fig4}
%fig4(231030).pdf
Phase diagram for $\theta=24^{\circ}$.
The left side panel: The allowed region denoted by the vertical gray band with  $\pm H^{\rm orb}_{\rm c2}(\theta=24^{\circ})$.
$H_{\rm eff}(\theta=24^{\circ})$ is showed by the red line together with those with $H\parallel b$ and $\theta=45^{\circ}$.
Above $H_{\rm m}$ it enters the allowed region up to $H^{\rm AUL}_{\rm c2}(\rm upper)$. The upper bound of the A$_2$ phase as shown in the
right hand panel is limited by the region $\gamma>\gamma_{\rm N}$.
In the middle fields, the A$_2$ phase reappears above $H_{\rm rot}$.
We note that $H_{\rm rot}(\theta=24^{\circ})=16{\rm T}/\cos24^{\circ}=17.5$T yields
the tetra-critical point at that field.}
\end{figure}

We note here that the combination between $H_{\rm c2}^{\rm AUL}(\theta)$
and the realized field region for  $\gamma(H)>\gamma_{\rm N}$ determines the
high field $A_2$ phase.
As shown in Fig.~\ref{fig4} for $H\parallel (011)$-direction ($\theta\sim23.8^{\circ}$)
the phase diagram is most intricate where the MSC phase and HSC phase coexist.
As seen from Fig.~\ref{fig4}, $H_{\rm eff}(H)$ denoted by the red line is within the allowed region in 
between $H_{\rm rot}<H_{\rm eff}<H_{\rm c2}^{\rm AUL}\sim$30T below $H_{\rm m}$.
Then it reappears above $H_{\rm m}$ up to the field for the extra DOS to be available.
Note that $H_{\rm c2}^{\rm AUL}(\rm upper)$ is far above, but it is not realized.

\begin{figure}
\begin{center}
\includegraphics[width=4.5cm]{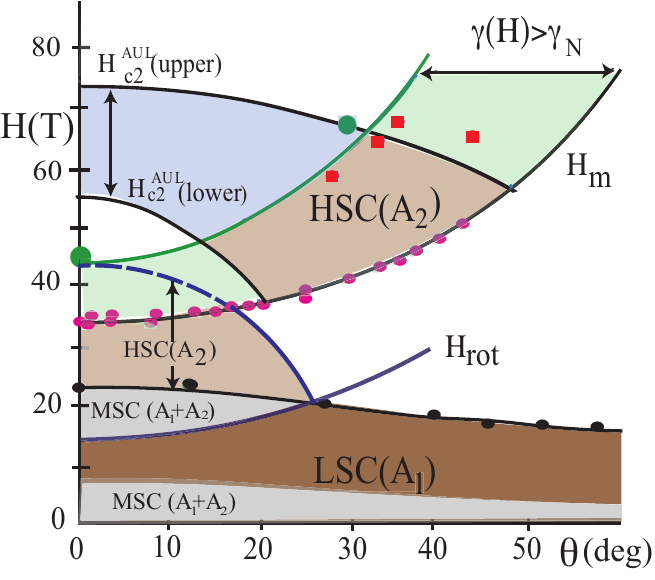}
\end{center}
\caption{
%fig5(231011).pdf
The phase diagram in the $H$-$\theta$ plane.
The gray band is defined by the upper and lower $H_{\rm c2}^{\rm AUL}$.
The green band is defined by the region for $\gamma(H)>\gamma_{\rm N}$.
In the overlapping region of the two bands HSC(A$_2$) phase appears, yielding the
highest $H_{\rm c2}$ around $\theta=35^{\circ}$. 
In the low angle  $\theta$ region HSC(A$_2$) phase appears below $H_{\rm m}$
below which MSC(A$_2$+A$_2$) exists. Inside MSC(A$_1$) in lower field bottom region
MSC(A$_2$+A$_2$) appears. $H_{\rm m}$ and $H_{\rm rot}$ behave like $\propto 1/\cos(\theta)$.
The red squares are the superconducting transitions~\cite{ran2} . 
The round green dots in $\theta=0^{\circ}$ and 28$^{\circ}$ are the upper limit of 
the observed points~\cite{miyake,miyake2} at which $\gamma(H)=\gamma_{\rm N}$.
\label{fig5} 
}
\end{figure}

To understand how $H_{\rm c2}(\theta)$ becomes maximal around $\theta\sim40^{\circ}$,
we summarize various lines which constrain the realized $H_{\rm c2}$ in Fig.~\ref{fig5}.
As seen from it, LSC containing the $A_1+A_2$ phase at the bottom occupies the lowest field region
and varies as a function of $\theta$ according to the Fermi velocity or mass anisotropy.
MSC appears in the middle fields above $H_{\rm rot}$, which forms the tetra-critical point
in the $H$-$T$ phase diagram for a given $\theta$. This is followed by HSC($A_2$)
appearing just above  $H_{\rm rot}$ and below $H_{\rm m}$.
As $H_{\rm rot}(\theta)$ becomes higher as $\theta$ increases
and eventually leaves the LSC region and also $H_{\rm c2}^{\rm AUL}$ sharply decreases, 
MSC and HSC cease to exist below $H_{\rm m}$.
This happens at around $\theta\sim30^{\circ}$.
Near this angle, the higher HSC($A_2$) appears above $H_{\rm m}$, whose region is sandwiched
by the band defined by $H_{\rm c2}^{\rm AUL}(\rm upper)$ and by $H_{\rm c2}^{\rm AUL}(\rm lower)$,
and the other band defined by the region $\gamma(H)>\gamma_{\rm N}$. 
The overlapping field region of the two bands above $H_{\rm m}$ denoted by HSC(A$_2$) in Fig.~\ref{fig5}
gives rise to the highest upper critical field.
It is understood that the coexisting two HSCs below and above $H_{\rm m}$ are rather special 
and only appear in the limited $\theta\sim24^{\circ}$.
Generically the $H$-$T$ phase diagram belong to either that in $H\parallel b$-axis shown in Fig.~\ref{fig2}(c)
or that shown in Fig.~\ref{fig3}.

In connection with the phase diagrams above, we study comparatively the cases in URhGe in Appendix B, and discuss  
the phase diagrams consisting with multiple phases under pressure to strengthen our
idea in Appendix C.

\subsection{High field phases in $bc \rightarrow a$}

It is interesting to examine how the high field phase HSC(A$_2$) of the $bc$-plane displayed in Fig.~\ref{fig5} evolves
when the field direction is tilted from this plane toward the $a$-axis because our theory can be checked further. 
This is experimentally done by Lewin et al~\cite{lewin}.
They show further higher $H_{\rm c2}$ surpassing 70T or more limited still experimentally. 
In the following we keep the same set of the parameters and the procedures used so far.
To construct the phase diagrams for $H$ tilted from the $bc$-plane toward the $a$-axis by the angle $\phi$,
we need to know magnetization curves for $M_b(H)$ for arbitrary angle $\phi$ first.

\begin{figure}
\begin{center}
\includegraphics[width=8cm]{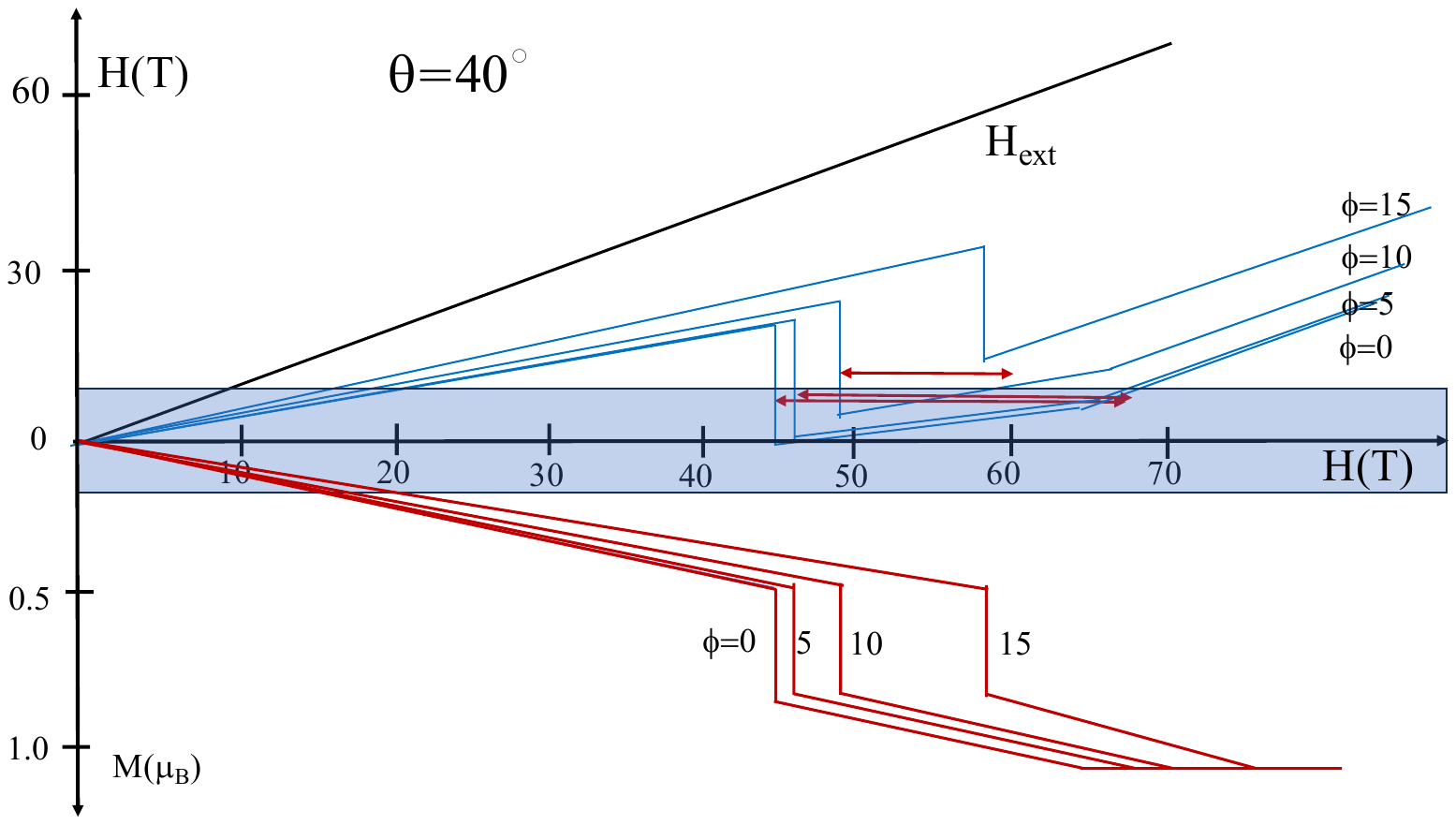}
\end{center}
\caption{
Magnetization curves (red curves) of the $M_b(H)$ component projected onto the applied field direction for the 
selected angles $\phi$ measured from the $bc$-plane under the fixed angle $\theta=40^{\circ}$.
The corresponding blue curves indicate the effective fields relative to the external field $H_{\rm ext}$. The allowed region
is shown by the gray band along the $H(T)$ axis. The double sided red arrows show the HSC(A$_2$) allowed.
\label{halo40} 
}
\end{figure}

The magnetization component $M_b(H)$ projected onto the field direction is obtained by considering the fact 
that all the meta-magnetic transition fields $H_{\rm m}$ are scaled and collapsed into
a single curve independent of the angle $\theta$, or all the $H_{\rm m}(\phi)$ curves 
for various $\theta$ are scaled to lead to
the identical curve.
This remarkable independence termed as ``orthogonality'' 
by Lewin et al~\cite{lewin} allows us to reproduce $M_b(H)$ for arbitrary angle $\phi$.
Namely starting with the known magnetization curve $M_b(H)$ for $H$$\parallel$$b$-axis,
the $M_b(H)$ component and the jump of $M_b(H)$ for the angle $\phi$  are obtained 
by shifting $H_{\rm m}(\phi)$ and by taking the projection onto the field direction.
The resulting $M_b(H)$ components for various $\phi$ values in $\theta =40^{\circ}$, for example,
 are shown by
the red curves in Fig.~\ref{halo40}.
These lead to the effective fields $H_{\rm eff}$ displayed by the blue curves for the corresponding $\phi$ values.
Thus the curves are within the allowed region for HSC to exist, indicated by the double sided arrows there.

\begin{figure}
\begin{center}
\includegraphics[width=9cm]{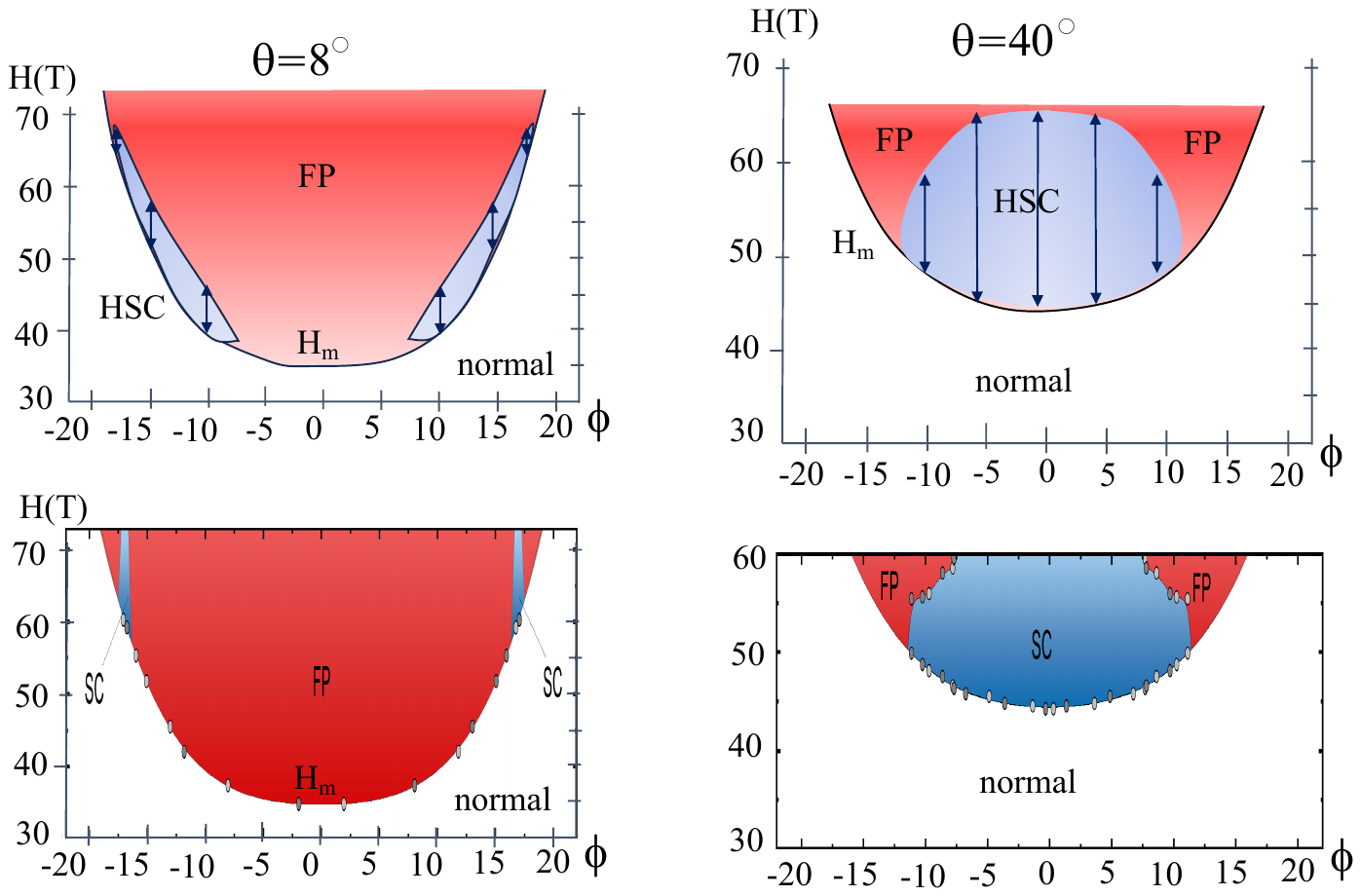}
\end{center}
\caption{
The resultant HSC regions  as a function of $\phi$ for the $\theta$=$8^{\circ}$ and $40^{\circ}$ cases (upper panel).
The experimental data for the two cases are referred to by Lewin et al~\cite{lewin} (lower panel).
For the case of $\theta$=$40^{\circ}$  the HSC appears at $\phi$=0 above $H_{\rm m}$ while it is absent for $\theta$=$8^{\circ}$.
This corresponds to these in Fig.~\ref{fig5}.
\label{halor} 
}
\end{figure}

According to Fig.~\ref{halo40}, we draw the phase diagram for $\phi$=40$^{\circ}$ in the top panel in right hand side in Fig.~\ref{halor}.
It is seen from Fig.~\ref{halor} that above $H_{\rm m}(\phi)$ the HSC region appears from $\phi$=0 up to
$\phi\sim$=10 and quickly disappear as $\phi$ increases, which is compared with the experimental data shown in the bottom panels by 
Lewin et al~\cite{lewin}.
Similarly, we can evaluate the case for $\theta=8^{\circ}$ by the same way mentioned
whose result is shown in the left column of Fig.~\ref{halor}.
It is seen that starting with $\phi$=0 where the HSC is absent, by increasing the angle $\phi$
the HSC gradually appears and fades way. In the corresponding experimental phase diagram shown
in the bottom panel is seen barely the tiny  HSC regions in high $\phi$ angles.
The evolution of the HSC in the $bc$-plane toward the $a$-axis is satisfactorily reproduced in a qualitative level.
This is another demonstration of the correctness of our theory.
.pdf

\begin{figure}
\begin{center}
\includegraphics[width=6cm]{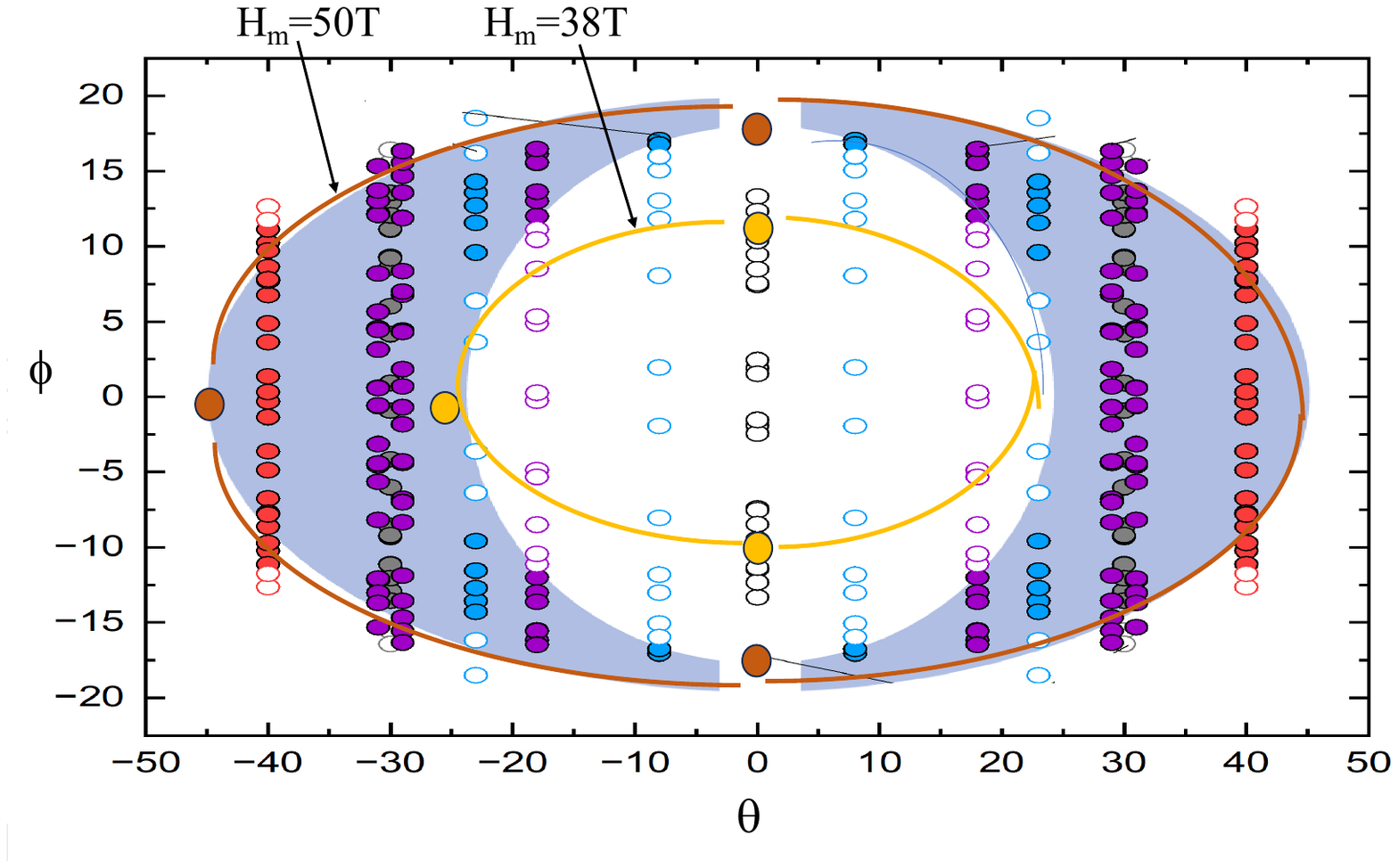}
\end{center}
\caption{
The experimental data points in the $\theta$-$\phi$ plane 
which correspond to the HSC to exist indicated by the filled circles~\cite{lewin}.
The grey regions indicate the HSC regions projected onto the $\theta$-$\phi$ plane.
Between the two contours of $H_{\rm m}$=38T and 50T there exist the HSC phases
 according to the present theory.
\label{contour} 
}
\end{figure}

These results can be easily understood physically and intuitively as follows:
As shown in Fig.~\ref{contour} where the contour map of $H_{\rm m}(\theta,\phi)$ is displayed as the functions of 
 $\theta$ and $\phi$, together with the experimental data for the HSC to exist~\cite{lewin} in this plane.
 The equi-$H_{\rm m}$ contour forms as an elliptic curve obtained by combining the two data sets
 of the $b$-axis$\rightarrow$$c$-axis results~\cite{miyake} and the $bc$-plane$\rightarrow$$a$-axis results~\cite{lewin}.
 The region sandwiched by $H_{\rm m}(\theta,\phi)$=38T and 50T is for the HSC to exist.
 For example, along the $\phi$=0 axis we have seen from Fig.~\ref{fig5} that the HSC exists
 from $H$=38T up to $H$$\sim$50T occurring at $\theta\sim45^{\circ}$ indicated by the HSC(A$_2$).
 This squared-like region of the HSC in Fig.~\ref{fig5} moves up to a higher field when $H$ is tilted toward the 
 $a$-axis. Thus in Fig.~\ref{contour} the HSC corresponds to the grey regions in between the two contours
 as shown by the experimental data with the filled points.
 It is also understood from Fig.~\ref{fig5} that the maximum $H_{\rm c2}$ of UTe$_2$ under ambient pressure
is less than or nearly equal to 70T throughout the ($\theta$, $\phi$)-plane. There exists no higher 
$H_{\rm c2}$ hidden in UTe$_2$.

\section{Knight shift}

\subsection{KS for $H\parallel a$-axis}

The Knight shift for $H\parallel a$-axis decreases below $T_{\rm c}$ for high-quality samples~\cite{},
which is contrasted with that for $T_{\rm c}$=1.6K samples without KS decreasing~\cite{ishida4}.
We now discuss this phenomenon in terms of our nonunitary scenario. At first sight, it is at odds with it
because the Cooper pair polarization $\boldsymbol S \parallel a$-axis at low $H$.
We solve this puzzle in the following.

We first note that the absolute value of $|\Delta K_a|$ for the $a$-axis is abnormally large~\cite{matsumura}, 
namely $|\Delta K|= 3.7\sim4.2\chi_{\rm N}$ as described in Appendix E.
To understand the negative and large $|\Delta K_a|$ 
we examine the GL coupling term $\kappa {\boldsymbol M}_a\cdot {\boldsymbol \eta}\times {\boldsymbol \eta}^*$
between the magnetization along the $a$-axis and the d-vector.
Under the quantization axis parallel to the $a$-axis, it can be written as
 
%$$\kappa {\boldsymbol M}_a\cdot {\boldsymbol \eta}\times {\boldsymbol \eta}^*=\kappa M_a(\Delta^2_{\uparrow}-\Delta^2_{\downarrow}).$$

\begin{eqnarray}
\kappa {\boldsymbol M}_a\cdot {\boldsymbol \eta}\times {\boldsymbol \eta}^*=\kappa M_a(\Delta^2_{\uparrow}-\Delta^2_{\downarrow}).
\label{Delta}
\end{eqnarray}

\noindent
Here we employ the notations $\Delta_{\uparrow}$ and $\Delta_{\downarrow}$
to explicitly express the spin moment direction.
If $\kappa<0$, which we consider in the following, then the split transition
temperatures $T_{\rm c1}$ and $T_{\rm c2}$ are identified  $T_{\rm c1}=T_{c\downarrow}$
and $T_{\rm c2}=T_{c\uparrow}$ respectively.
Namely upon lowering $T$, $\Delta_{\downarrow}$ is condensed first and
further lowering $T$, $\Delta_{\uparrow}$ is followed at $T_{\rm c2}$.
Since the Cooper pair spin moment $\boldsymbol S$ defined by  Eq. (\ref{spin})
is locked by the magnetization direction $\boldsymbol M_a$ through the above GL coupling,
not the external field, independent of the positive or negative direction of the external field $\boldsymbol H$, 
$\Delta_{\downarrow}$ appears below $T_{\rm c1}$ because it is locked by $\boldsymbol M_a$.
This situation is contrasted in the conventional case where $\Delta_{\uparrow}$ always appears first 
to save the Zeeman energy
even if the external field direction is reversed, leading to the
paramagnetic response $\chi_s=+\chi_{\rm N}>0$.
Thus it immediately leads to $\Delta K_a<0$ below $T_{\rm c1}$.
The A$_1$ phase with $\Delta_{\downarrow}$ responds diamagnetically to the external field.

Let us consider the magnitude of $|\Delta K_a|$.
The spin susceptibility $\chi_a(T)$ near $T_{\rm c1}$ is calculated starting with the GL functional
given by Eq. (\ref{bulk2}), following the procedure by Takagi~\cite{takagi} on the $^3$He-A phase 
under applied fields,

\begin{eqnarray}
f=a_0 (T-T_{\rm c1}(H))\Delta^2_{\downarrow}
\label{chi}
\end{eqnarray}

\noindent
with $a_0=N(0)/T_{\rm c1}$ and $T_{\rm c1}=T_{\rm c0}+{\kappa\over a_0}M_a$.
The magnetization change $\delta M$ is derived as 

\begin{eqnarray}
\delta M=-{\partial f\over \partial H}=a_0 \kappa{\partial M\over \partial H}\Delta^2_{\downarrow}.
\label{chi}
\end{eqnarray}

\noindent
This leads to 

\begin{eqnarray}
{\delta M\over M_{\rm Pauli}}={\Delta^2_{\downarrow}\over2\mu_{\rm B} HT_{\rm c1}}{\kappa\over a_0}{\partial M\over \partial \mu_{\rm B}H}.
\label{chi}
\end{eqnarray}

\noindent
Substituting the weak coupling BCS value for $\Delta_{\downarrow}=1.75T_{\rm c1}$,
${\kappa\over a_0}=2.7$K/$\mu_{\rm B}$ and ${\partial M\over \partial \mu_{\rm B}H}=0.07\mu_{\rm B}$/K
at $H$=1T~\cite{miyake}, we obtain ${\delta M/ M_{\rm Pauli}}=1.3$, which is smaller than the estimated observe value
3.7$\sim$4.2. If taking into account the strong coupling value $\Delta_{\downarrow}=(1.5\sim2.0)\times1.75T_{\rm c1}$,
we find ${\delta M/M_{\rm Pauli}}=2.9\sim 5.2$, agreeing with the experimental estimate.
Note that the strong coupling effect is evident from the large specific heat jump~\cite{aokidHvA} 
$\Delta C/\gamma_{\rm N}T_{\rm c1}\sim 2.7$ 
with the $T_{\rm c}$=2.1K samples compared with 1.43 in BCS.
This agreement implies that the mechanism of the KS along the $a$-axis is quite different from the
usual case for $\boldsymbol H\perp \boldsymbol S$. This mechanism is somewhat similar to that in $^3$He-A$_1$~\cite{takagi}.
In Appendix C we estimate the KS value different from this due to Miyake~\cite{miyakeriron}, 
which yields a similar value.

\begin{figure}
\begin{center}
\includegraphics[width=8cm]{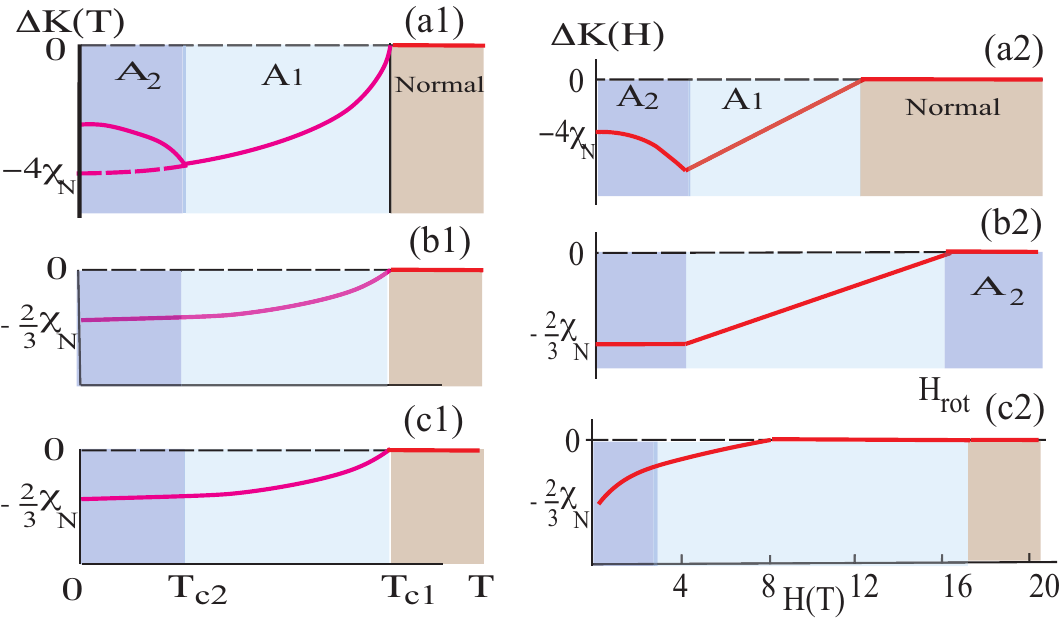}
\end{center}
\caption{\label{fig11}
%chi(231020).pdf.
The Knight shift $\Delta$K or the spin susceptibilities in the superconducting state
as functions of $T$ (left column) at low $H$ and $H$ (right column) for three field orientations.
Upon lowering $T$, $\Delta$K$(T)$ strongly decreases below $T_{\rm c1}$ for $H\parallel a$
and increases below $T_{\rm c2}$ when the A$_2$ appears (a1) while $\Delta$K$(T)$
behaves as in the Yosida function for $b$ (b1) and $c$ (c1) axes.
$\Delta$K$(H\parallel a)$ decreases in the A$_2$ state and increase when entering the A$_1$ state
with increasing $H$ (a2). $\Delta$K$(H\parallel b)$ remains a constant at lower $H$ and increases
when entering the A$_1$ state with increasing $H$ (b2). $\Delta$K$(H\parallel c)$ keeps
gradually increasing and reaches $\Delta$K=0 around 8T (c2).
}
\end{figure}

As shown in Fig.~\ref{fig11}(a1), with decreasing $T$ the $T$-dependence of  $\Delta K(T)$ at low $H$
exhibits a  large drop below $T_{\rm c1}$. Below $T_{\rm c2}$, or $T_{{\rm c}\uparrow}$ the decrease stops and
slightly increases because $\Delta_{\uparrow}$ starts appearing as shown in Fig.~\ref{fig2}(d).
It is seen from Fig.~\ref{fig11}(a2) that with increasing $H$,
$\Delta K(H)$ at low $T$ shows a decrease toward
$H_{\rm c2\uparrow}$ at which $\Delta_{\uparrow}=0$ as shown in Fig.~\ref{fig2}(d). Then it increases up to
$H^a_{\rm c2}$=12T toward the normal state because in this field region
$\Delta_{\downarrow}$ diminishes toward it.

\subsection{KS for $H\parallel b$-axis}

As shown in Fig.~\ref{fig11}(b1), with decreasing $T$  the $T$-dependence of  $\Delta K(T)$ at low $H$
for $H\parallel b$-axis exhibits a drop below $T_{\rm c1}$ because $\boldsymbol S$ points to the $a$-axis
or one of the two d-vector components is parallel to the field direction. 
Thus $\Delta K(T\rightarrow 0)=-{2\over 3}\chi_{\rm N}$. The overall $T$-dependence follows the
usual Yosida function~\cite{tinkham}. 
We notice that the actual observed value of $\Delta K(T\rightarrow 0)$ is quite small~\cite{ishida1,ishida2,ishida3}
 compared with this
ideal one due to the unknown reason characteristic to heavy Fermion supercoductors.

As for the $H$-dependence of $\Delta K(H)$ at low $T$ is displayed in Fig.~\ref{fig11}(b2).
At the lowest $H$, $\Delta K(H)$ starts with $-{2\over 3}\chi_{\rm N}$.
Corresponding to the d-vector rotation phenomenon $\Delta K(H)$ gradually increases toward the
normal state value. When the d-vector rotation is completed at $H_{\rm rot}=15$T, the polarization vector
$\boldsymbol S$ becomes perfectly parallel to the $b$-axis. $\Delta K(H)$=0 so that the spin susceptibility $\chi_{\rm s}$ 
returns to $\chi_{\rm N}$~\cite{ishida5}.

\subsection{KS for $H\parallel c$-axis}

As shown in Fig.~\ref{fig11}(c1), with decreasing $T$ the $T$-dependence of  $\Delta K(T)$ at low $H$
for $H\parallel c$-axis exhibits a drop below $T_{\rm c1}$ in the A$_1$ phase.
Since $\boldsymbol S$ points to the $a$-axis,
$\Delta K(T\rightarrow 0)=-{2\over 3}\chi_{\rm N}$ for this field orientation.
The functional form of $\Delta K(T)$ is described by the Yosida function.

The field dependence of $\Delta K(H)$ at low $T$ is shown in Fig.~\ref{fig11}(c2).
According to the experiments~\cite{ishida1,ishida2,ishida3}, $\Delta K(H)$ increases at the lowest $H$ and locks to the
normal state value around 8T, which is smaller than that for the $b$-axis, indicating that
the locking energy to the crystalline lattice is weaker in this direction.
This means that the original $SO(3)^{\rm spin}$ is weakly broken due to the spin-orbit coupling.

\subsection{KS under pressure}

It is noteworthy to interpret the KS experiments under pressure $P=1.2$GPa by Kinjo et al~\cite{kinjo} 
for $H\parallel b$-axis to check the overall consistency of our scenario. 
According to them, as lowering $T$, $\Delta K(T)$ is unchanged below $T_{\rm c1}$ (see Fig.~\ref{fig9}(b3) in Appendix C) where
they measure KS along the $T$ axis).
Upon further lowering $T$, it starts decreasing below $T_{\rm c2}$, and simultaneously the resonance width is broadened. At first sight, it seems unphysical: The low $T$ phase deliberately loses
the Zeeman energy since the high $T$ phase is the most stable one by attaining  $\boldsymbol S$ parallel to the $b$-axis.

This paradox can be understood as follows:
Since the A$_1$ phase has the spin polarization ${\bf S}\parallel b$-axis, $\Delta K(T)$ 
should be unchanged below $T_{\rm c1}$ as observed. 
Thus the A$_1$ phase as the high $T$ state has the order parameter $\Delta_{\uparrow}$
where the spin quantization axis must be the $b$-axis.
Upon entering the A$_2$ phase as the low $T$ state whose spin polarization is opposite to it below $T_{\rm c2}$.
Namely the A$_2$ phase has the order parameter $\Delta_{\downarrow}$.
$\Delta K(T)$ must decrease 
because the spin polarization $\boldsymbol S$ points to the antiparallel direction relative to the applied field.
This situation is exactly same as  in the $H\parallel a$-axis case at the ambient pressure. 
The concomitant resonance width broadening is fully consistent with the spin texture formation in the mixture of the 
 A$_1$ and  A$_2$ phases~\cite{tsutsumi}.
 
 Thus the results are important in the following points:\\
 \noindent
 (1) The high $T$ phase of  A$_1$ is originally inherited from the A$_2$ phase under the ambient pressure
 where the spin polarization points to the opposite direction of the applied field as mentioned above.
 As pressure increases, this phase becomes the low $T$ phase after passing the critical pressure $P_{\rm cr}$=0.14GPa
 at which two phases cross each other~\cite{pressure1}. 
 Thus above $P_{\rm cr}$ the high $T$ phase has the spin polarization ${\bf S}$
 directed to the applied field.\\
 \noindent
(2)The resonance width in A$_1$ is narrow because all spin polarizations for Copper pairs are parallel.
Since the low temperature phase A$_2$ is mixed by the A$_1$ phase, the spin polarization forms a spin texture,
thus making the resonance width wider~\cite{tsutsumi}.

\section{Discussion}

\subsection{Spin-singlet scenario}

%see Note 23.8.31
It may be instructive to examine the possibility of the spin-singlet scenario
from both positive and negative sides.\\
\noindent
(1) The observed $H_{\rm c2}$ for all three principal directions exceeds by far
the Pauli paramagnetic limiting field. This difficulty for the spin-singlet scenario
may be circumvented by considering the cancelation field due to induced localized moment.
This avoiding Pauli limiting effect was discussed successfully in connection with CeRh$_2$As$_2$~\cite{machida4}.\\
\noindent
(2) The KS decreases for all three principal directions are obviously explained by a spin-singlet pairing.
However, it should be noted that the KS values are not equal expected for the spin-singlet scenario.\\
\noindent
(3) Although the time reversal symmetry either preserved or broken is under debate~\cite{kaptulnik,madhavan,sonier1},
it is easy to imagine that the simplest spin-singlet  $s$-pairing is realized when the former is the case. \\
\noindent
(4) The gap symmetry is also under discussion whether it is a nodal gap~\cite{metz,kittaka,shibauchi} or a full gap~\cite{matsuda}.
If the latter is true, the simplest spin-singlet  $s$-pairing is most possible.\\
\noindent
(5) There are reports~\cite{ran,sonier} to support ferromagnetic fluctuations which are believed to be favorable
for the spin-triplet pairing formation, making the spin-triplet scenario less possible. 
The reported antiferromagnetic fluctuations~\cite{af1,af2,af3} and incommensurate spin-density-wave long-range order~\cite{knafo2}
under pressure may be favorable for higher angular momentum
pairing other than the simple s-wave one.\\
\noindent
(6) The multiple phase diagrams under ambient and applied pressures are most difficult to explain
in terms of the spin-singlet scenario unless accidentally different phases are nearly degenerate.
Generically the spin-triplet pairing has various degeneracies in its spin space.
In particular within our weak-spin orbit case the SO(3)$^{\rm spin}$ has many degenerate
phases because of the rich internal degrees 3$\times$2 of spin freedom.

\subsection{DOS mystery}

\begin{figure}
\begin{center}
\includegraphics[width=6cm]{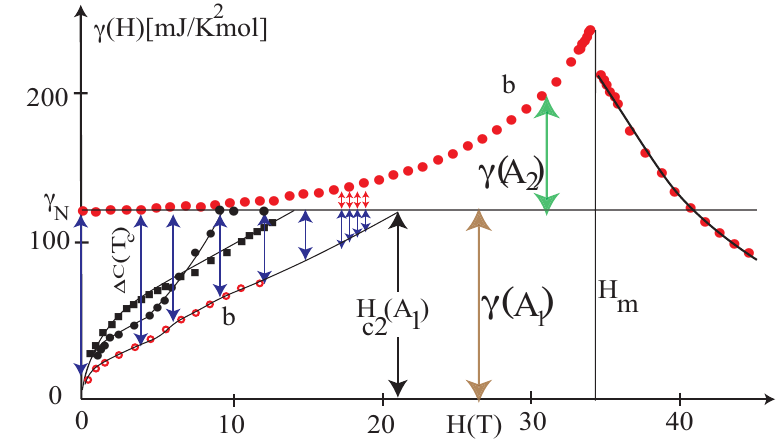}
\end{center}
\caption{\label{fig12}
%fig10(231102).pdf.
$\gamma(H)$ in the normal state indicated by the filled red dots for $H\parallel b$-axis~\cite{miyake}.
$\gamma(H)$ in the superconducting state~\cite{roman} indicated by the empty red dots and its extrapolation
by the black curve. It reaches $\gamma_{\rm N}$ at 20T at which the A$_1$ state disappears.
The extra DOS indicated green arrow denoted by $\gamma(A_2)$ coming from the enhanced DOS
while DOS $\gamma(A_1)$ for the A$_1$ phase occupies the whole  $\gamma_{\rm N}$.
The black  and red arrows denoted by $\Delta C(T_{\rm c})$ show the $\gamma(H)$ estimated by the
specific heat jump data.
}
\end{figure}

We show the DOS $\gamma(H)$ evolutions~\cite{miyake,miyake2,roman} under fields in Fig.~\ref{fig12}.
For $H\parallel b$-axis, the normal state $\gamma(H)$ is displayed by the filled red dots
and the empty red dots are in the superconducting state.
It is seen from this that the SC $\gamma(H\parallel b)$ reaches the normal DOS $\gamma_{\rm N}$
around $H\sim$21T by extrapolating linearly to higher $H$ where the A$_1$ phase ends, denoted by $H_{\rm c2}(A_1)$.
Above this field, the A$_2$ becomes reappearing (see Figs.~\ref{fig2}(a) and (c)). 
Thus the DOS for the A$_2$ must be supplied by some
subsystems other than the conduction electron subsystem.
Since the normal state DOS denoted by the filled red dots increases 
beyond $\gamma_{\rm N}=120$ mJ/K$^2$mol which is supplied from the magnetic subsystem,
this extra-DOS indicated by $\gamma(A_2)$ (the green arrow in Fig.~\ref{fig12})
now available for the A$_2$ phase to reappear above $H$=21T.
Otherwise, it is impossible to reappear because the normal DOS $\gamma_{\rm N}$ is exhausted by the
A$_1$ phase when $H$ reaches $H_{\rm c2}(A_1)$.

As denoted by the red arrows in Fig.~\ref{fig12} estimated by the specific heat jumps
under fields~\cite{rosuel}  the DOS for the A$_2$ phase exactly corresponds to these extra-DOS supplied by the magnetic subsystem (see Appendix D for details on how to evaluate those arrows).
Thus it is clear that the A$_2$ phase only exists near $H_{\rm m}$ where the enhanced DOS 
beyond $\gamma_{\rm N}$ is available.

As shown in Fig.~\ref{fig10} in Appendix D,  the enhanced DOS region beyond $\gamma_{\rm N}$
is much wider for $\theta=28^{\circ}$ than that for $\theta=0^{\circ}$, namely $H\parallel b$-axis.
We denote this field region $\gamma(H)>\gamma_{\rm N}$ as a function of $\theta$ in 
Fig.~\ref{fig5}. Within this band, the A$_2$ phase can exist. This is a necessary
condition for the A$_2$ phase to reappear above $H>H_{\rm m}$.
Therefore, in principle the A$_2$ phase can persist along $H>H_{\rm m}(\theta)$,
but in UTe$_2$ another necessary condition coming from $H^{\rm AUL}_{\rm c2}$
prevents  it from attaining the higher $H_{\rm c2}$.
On the other hand, in URhGe the reentrance A$_2$ phase persists up to
higher angles $\varphi$ measured from the $b$-axis as seen from Figs.~\ref{fig7} and ~\ref{fig8}
in Appendix C.

Here there is a mystery that is not solved:
there is no available DOS for the low field A$_2$ phase expected below a few T
since the available normal DOS is exhausted by the A$_1$ phase.
One possibility to resolve it is to assume that the low field A$_2$ phase
uses a small portion of the normal DOS. If it is true, we may understand that
at $T_{\rm c2}$ there is no detectable specific heat jump.

\subsection{Hints for the second transition $T_{\rm c2}$ at $H$=0 under ambient pressure}

\subsubsection{T$_1(T)$ anomaly}
According to Matsumura, et al~\cite{matsumura2},
1/T$_1(T)$ on $^{125}$Te-NMR under $H\parallel b$-axis
exhibits an anomalous rise below 0.5K for $H$=0.35T and 0.65T, but not for
3T and 5T which show a monotonous decrease upon further lowering $T$.
These anomalous rises in 1/T$_1(T)$ suggest the appearance of the additional 
DOS corresponding to $\sim$0.2$\gamma_{\rm N}$~\cite{matsumura2}.
Thus we may regard it as $T_{\rm c2}$=0.5K since the building-up extra-DOS 
comes from the second transition to newly form of the gap edge singularity at $E=\pm \Delta_{\uparrow}$.
This experiment~\cite{matsumura2} also suggests that the small pocket region in the $H$-$T$ phase diagram
vanishes around $H_{\rm c2}(A_2)$=3T$\sim$5T as indicated in Fig.~\ref{fig2}(d).

\subsubsection{$H_{\rm c1}(T)$ anomaly}

It is seen from the data~\cite{shibauchi2} 
that the $T$ dependences of $H_{\rm c1}$ exhibit anomalous upward changes for all
field directions, in particular, $H\parallel c$-axis is eminent.
Naive extrapolation of this anomaly may start around $T_{\rm c2}$=0.5K.
This kind of H$_{\rm c1}(T)$ anomalies is also observed by Paulsen et al~\cite{paulsen}.
Since the applied fields are around a few mT, those measurements suggest
that the phase transition at $T_{\rm c2}$=0.5K may occur at $H$=0.

\subsubsection{$\gamma(H)$ anomalies}

As displayed in Figs.~\ref{fig12} and~\ref{fig10}, it is seen that all $\gamma(H)$ curves in the
SC state~\cite{roman} exhibit anomalies around a few T. Especially $\gamma(H\parallel a)$
is eminent with a sharp kink at 4T. Those anomalies are consistent with the idea that
the small pocket regions for the A$_2$ phase exist in the $H$-$T$ phase diagrams
with $H_{\rm c2}(A_2)$=$4\sim 6$T, depending on the field orientations as displayed in Figs.~\ref{fig2}(c) and (d).

\section{Conclusion and summary}

\subsubsection{Conclusion}

We have extended our theoretical efforts toward a complete understanding of
the pairing symmetry in UTe$_2$.
In particular, we have refined our construction work of the peculiar phase diagram for $H\parallel$$b$-axis.
The present work is motivated by the recent several remarkable experimental progresses
using high-quality samples with $T_{\rm c}$=2.1K, including the newly found fourth internal line~\cite{sakai}
with an almost horizontal slope going into the tetra-critical point in the $H\parallel b$-axis
together with the positive sloped $H_{\rm c2}$ departing from it.  
This improbed procedure to reconstruct the $b$-axis phase diagram leads us to obtain 
the off-axis phase diagrams with a record-high $H_{\rm c2}$ in a consistent manner.
These remarkable phase diagrams are successfully reproduced here by assuming a nonunitary state
with equal spin pairs whose split transition temperatures are tuned by the underlying magnetization $M_b(H)$.
This theoretical framework also explains the Knight shift experiments consistently~\cite{matsumura,ishida1,ishida2,ishida3,ishida4,ishida5,kinjo}
by introducing the concept of the d-vector rotation~\cite{tsutsumi} occurring at the tetra-critical point.

Within this framework on the minimal assumptions we have accomplished the following
in this paper:\\

\noindent
(1) According to Sakai et al~\cite{sakai}, they identified the three phases LSC, MSC and 
HSC for the $H\parallel b$-axis. These phases are characterized in Table I. In addition to the three phases
the fourth phase LLSC as the lowest $H$ phase 
is proposed here. The Table I includes the charactor associated with the vortex core structures
in each phase, in particular whether it is pinned in crystal lattices or easy depinned, exhibiting the
flux flow under the applied current. It also decribes the width of the NMR resonance spectrum,
which matches with the observed characteristics~\cite{kinjo,matsumura}.\\
\noindent
(2) In addition to the peculiar phase diagram for $H\parallel b$-axis, we have shown that the
high field A$_2$ phase is reentrant  when the field orientation is tilted from the $b$-axis to the $c$-axis
and from the $bc$-plane toward the $a$-axis,
achieving extremely high $H_{\rm c2}$. This is caused by the screening effect of the external field by the
internal field due to the localized 5f moment $M_b$ along the $b$-axis, which exhibits the meta-magnetic transition.
This explains the truly multiple phases in this system.\\
\noindent
(3) We have pointed out similarities and differences from the sister materials URhGe in the phase diagram, 
maintaining the same scenario based on the nonunitary equal-spin pairing.\\
\noindent
(4) Under pressure the observed multiple phase is demonstrated to be connected to the multiple phase in
ambient pressure. This allows us to paint out the whole phase diagrams in a unified viewpoint.\\
\noindent
(5) We have analyzed the DOS data to show that the high field A$_2$ phase as HSC is needed for the
extra DOS, which is supplied from the localized magnetic subsystem. Thus for the A$_2$ phase
to appear the extra DOS associated with the meta-magnetic transition is indispensable, hinting at the
novel pairing mechanism.\\
\noindent
(6) We have explored thoroughly the KS experiments by NMR for all field orientations and under pressure
in a consistent manner and predicted the KS behaviors which are not done yet.\\

\begin{table}
  \caption{Characterizations of the four phases for $H\parallel b$}
  \label{table}
  \centering
  \begin{tabular}{lccccc}
 \hline  
%phase & $\chi_s/\chi_N$ &d-vector&singularity&Majorana&width*\\
phase & $\Delta K$ &d-vector&vortex core (flow)&width*\\
 \hline   \hline
 % LSC & 1/2 &d$\parallel$$B$ (uniform)&yes&core&wide\\
LLSC & decrease &spin texture&soft(flux flow)&wide\\
%\hline
% LSC & 1/2 &d$\parallel$$B$ (uniform)&yes&core&wide\\
 LSC & decrease&${\boldsymbol d}\parallel{\boldsymbol H}$ (uniform)&hard(pinned)&wide\\
%\hline
MSC & decreas & spin texture&soft(flux flow) & wide\\
%\hline
HSC & unchange &${\boldsymbol d}\perp{\boldsymbol H}$ (uniform)&hard(pinned)&narrow\\
\hline
 \end{tabular}
*width: NMR resonance width 
\end{table}

To strengthen our argumentation, we have to back it up in several aspects:\\
\noindent
(1) Since our arguments are qualitative or semi-quantitative,
we need more microscopic calculations based on a microscopic Hamiltonian.
This includes the $H_{\rm c2}$ evaluation quantitatively based on band calculations,
giving the realistic values of the anisotropy and accurate position of the tetra-critical point.\\
\noindent
(2) Microscopic computations are needed to understand the dynamical processes of the
formation and nucleation for spin textures to accurately describe the d-vector rotation
phenomena by extending our microscopic Eilenberger theory~\cite{tsutsumi} 
or Bogoliubov-de Gennes theory~\cite{isoshima}.\\
\noindent
(3) Although it is a notoriously difficult question and so far no one answers reasonably,
the origin of the pairing mechanism of this kind of particular superconducting state must be addressed.
Here we have tried to extract some hints related to this fundamental question which might give clues.
For example, as mentioned in Sec.VI in order to produce the high field A$_2$ phase above $H_{\rm m}$
the extra DOS must be supplied from the magnetic subsystem whose internal degrees of freedom associated
with the Kondo screening is liberated under strong external fields by more and more localizing the 5f moments.
On the other hand, the low field A$_2$ phase as LLSC expected to appear at $T_{\rm c2}$ under ambient pressure 
and $T_{\rm c1}$ under applied pressure seems to hold only tiny entropy compared with the A$_1$ phase
with holding the almost full DOS $\gamma_{\rm N}$=120mJ/mol K$^2$, not sharing half- and half-DOS for each.
It seems quite puzzling at this moment because the spin-down system $\Delta_{\downarrow}$ overwhelms the spin-up system
$\Delta_{\uparrow}$.\\
\noindent
(4) The pairing mechanism must be related to the underlying magnetism because the present system is governed
by the field-tuned 5f magnetic moment.
The recent discovery of the long-range
incommensurate spin-density-wave under high pressure~\cite{knafo2} and antiferromagnetic fluctuations~\cite{af1,af2,af3}
under ambient pressure may indicate a hint for the pairing glue. More broadly, the competing phase between superconductivity
and magnetism belongs to more general topics  not only in heavy Fermion materials~\cite{matsu,nokura,nakanishi,kato}, but also in high $T_{\rm c}$ 
cuprates~\cite{kivelson,keimer}.\\
\noindent
(5) In the present paper we do not touch upon the gap structure because it is not directly
relevant for our phase diagram construction. But its determination is essential to finally pin down
the proper pairing state because the spin part and orbital part of the order parameter must be mutually
consistent. At this moment there are conflicting opinions on the gap structure, either nodal or full gap.
To resolve it, we need further efforts both theoretically and experimentally.
Some of these tasks are underway.

Finally, we wish to list up the experimental works to be urgently performed:\\
\noindent
(A) According to the present scenario~\cite{tsutsumi}, the vortex core in the high field A$_2$ phase for $H\parallel b$-axis
contains a genuine spinless Majorana zero-energy mode. This can be detected by STM-STS via the local density of states measurements~\cite{ichioka}.\\
\noindent
(B) The spin textural changes associated with the d-vector rotation, whose precise field and temperature region is identified in
the present paper, can be also observed
by small angle neutron scattering (SANS) experiments.\\
\noindent
(C) The detailed $\gamma(H)$ measurements in higher fields for  the $H\parallel b$-axis
is necessary for clarifying the DOS mystery mentioned above.\\
\noindent
(D) The experimental check on whether or not the second transition at $T_{\rm c2}$=0.4$\sim$0.5K described in this paper
is needed by using high-quality samples with  $T_{\rm c}$=2.1K. The thermodynamic anomaly
is expected to be very small, thus high-resolution experiments are indispensable.
According to a recent paper~\cite{andersen}, it is pointed out that the second transition in a nonunitary state can be small
because of the presence of ${\boldsymbol d}\times{\boldsymbol d}^*$.\\
\noindent
(E) To precisely determine the gap structure as for the nodal topology, point or line, or full gap
and its reciprocal-space position, the angle-resolved specific heat~\cite{miranovic} or thermal-conductivity measurements
are highly desired.

\begin{figure}
\begin{center}
\includegraphics[width=6cm]{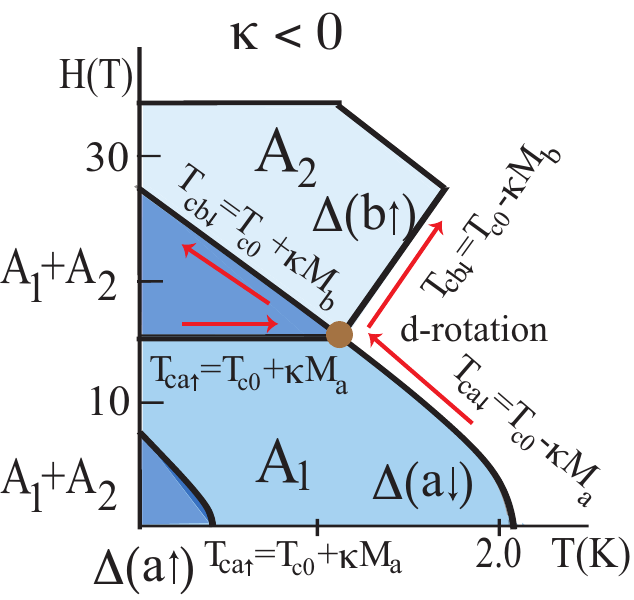}
\end{center}
\caption{\label{summary1}
The characterization of the order parameters and the transition lines for the phase diagram in $H$-$T$ plane for $H\parallel$$b$-axis.
}
\end{figure}

\begin{figure}
\begin{center}
\includegraphics[width=5cm]{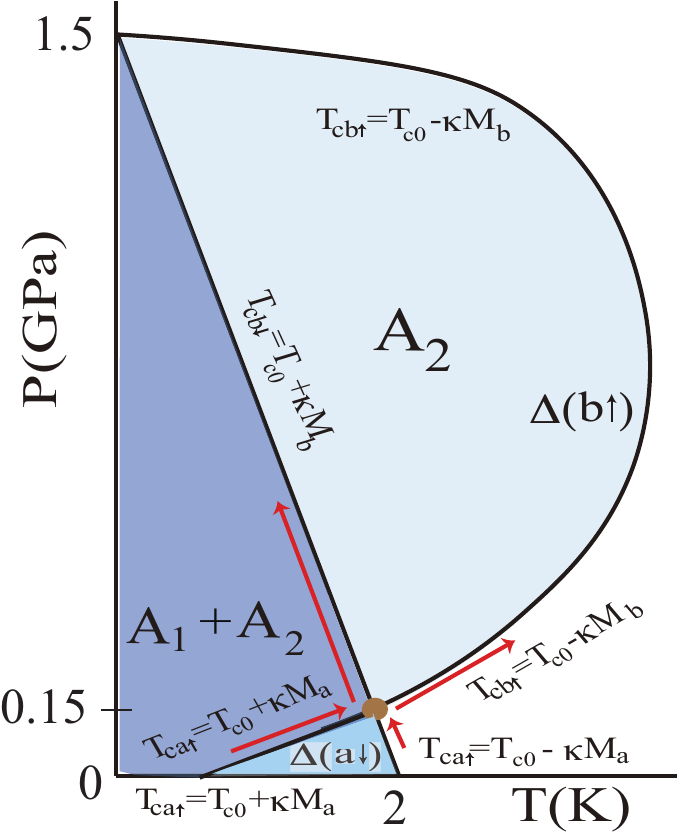}
\end{center}
\caption{\label{summary2}
The characterization of the order parameters and the transition lines for the phase diagram in $P$-$T$ plane.
}
\end{figure}

\subsubsection{Summary}
To facilitate future experiments,
it may be instructive to summalize the identified pairing states on the phase diagrams
for $H\parallel$$b$-axis and under pressure.
On the phase diagram for $H\parallel$$b$-axis, which is derived early in Figs.~\ref{fig1} and ~\ref{fig2}(c)
the relevant order parameters
are described in Fig.~\ref{summary1}.
Here we have introduced the notation for the order paramters, such as $\Delta(a\downarrow)$
where the quntization axis of the spin direction is specified and the transition temperatures
are specified, such as $T_{\rm ca\downarrow}=T_{\rm c0}-\kappa M_a$, indicating that $\Delta(a\downarrow)$
appears at this temperature governed by $M_a$.
The red arrows at the tetra-critical point denote the reconnections of the transition lines,
namely $T_{\rm ca\downarrow}=T_{\rm c0}-\kappa M_a$ at 2K
($T_{\rm ca\uparrow}=T_{\rm c0}+\kappa M_a$) reconnects to $T_{\rm cb\uparrow}=T_{\rm c0}-\kappa M_b$
($T_{\rm cb\downarrow}=T_{\rm c0}+\kappa M_b$) at TCP where the d-vector rotates.
Thus the low field phase $A_1$ (the high field phase $A_2$) is characterized by the order parameter $\Delta(a\downarrow)$
($\Delta(b\uparrow)$).

We notice that at TCP the four transitions meet together. They are all degenerate, allowing them to freely
reconnect so as to be maximally beneficially for them. In fact for $\Delta(b\uparrow)$ the transition line
$T_{\rm cb\uparrow}=T_{\rm c0}-\kappa M_b$ is most beneficially becasue the diamagnetic effect is absent
and the transition temperature can strongly increase as $M_b(H)$ grows toward higher fields in our case of $\kappa<0$.
These identifications are fully consistent with the Knight shift experiments.

Under pressure the $P$-$T$ phase diagram is shown in Fig.~\ref{summary2}.
The similarity between Figs.~\ref{summary1} and ~\ref{summary2} is noticable:
Namely the high pressure and high temperature phase corresponding to the $A_2$ phase
with $\Delta(b\uparrow)$ changes into the mixture phase of $A_1+A_2$ phase upon cooling 
where $\Delta(b\downarrow)$ appers at $T_{\rm cb\downarrow}=T_{\rm c0}+\kappa M_b$.
This identification of the spin structure of the order parameters is 
consistent  with the Knight shift experiment by Kinjo et al~\cite{kinjo} under $P=1.2$GPa.
At the tetra-critical point $P\sim$0.15GPa the d-vector rotation must occurs.

\acknowledgments
The author sincerely thanks K. Ishida, and S. Kitagawa for sharing unpublished results,
which motivates the present project. He is indebted to D. Aoki, and R. Movshovich
for showing their data prior to publication 
and T. Sakakibara, S. Kittaka, Y. Shimizu, A. Miyake, Y. Tokunaga, 
H. Sakai, M. Yamashita, T. Shibauchi, K. Hashimoto, K. Ishihara, and S. Imajo
for enlightening and fruitful discussions.
This work is supported by JSPS KAKENHI, Grant No. 21K03455 and No. 17K05553.
A part of this work was performed during stay at the Aspen Center for Physics, which is supported by National Science Foundation grant PHY-2210452.

\appendix

\section{Phase diagrams for $T_{\rm c}$=1.6K samples}

The samples with $T_{\rm c}$=1.6K exhibit no indication for the double transition at $T_{\rm c2}$
at the ambient pressure and $H$=0. Note that all the indications of the second transition $T_{\rm c2}$
in sec. VI C come from the samples with $T_{\rm c2}$=2K~\cite{matsumura,shibauchi2,roman}.
This is caused by the large residual DOS comparable to 0.5$\gamma_{\rm N}$,
which ultimately excludes the A$_2$ phase from realizing by pushing $T_{\rm c2}$ to a negative value.
We remark that the new samples with $T_{\rm c}$=2.1K have no
such residual DOS.
Other various eminent features discussed in the main text remain intact even 
in these samples with $T_{\rm c}$=1.6K. In particular, for $H\parallel b$-axis
unusual phenomena, such as 
the positive slope in $H_{\rm c2}$ above 15T, the successive transition under fields
and the high field reentrant SC for the field orientations along $\theta$ are observed in
the low-quality samples. Thus it is important to check whether or not the present theoretical
framework can explain these prominent features.

\subsection{Construction of the phase diagram for $H\parallel b$-axis}

\begin{figure}
\begin{center}
\includegraphics[width=5cm]{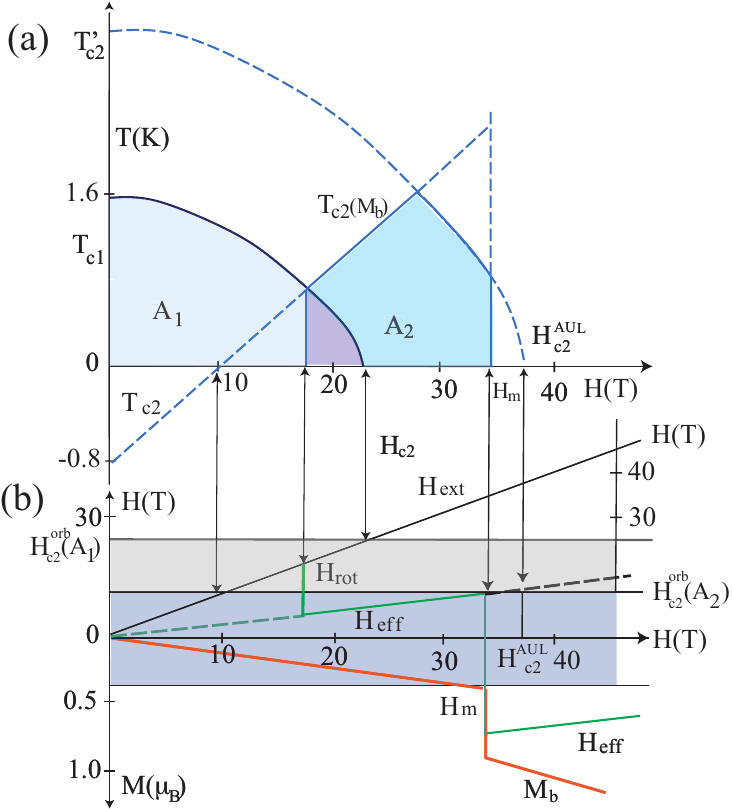}
\end{center}
\caption{\label{fig13}
%fig13(231102).pdf.
Construction of the phase diagram for $H\parallel b$-axis when $T_{\rm c1}<0$.
The lower panel shows the magnetization curve of $M_b(H)$ by the red line.
The green line is $H_{\rm eff}$ as a function of $H$ where the dotted lines are not realized.
The two horizontal bands exhibit the allowed regions $H^{\rm orb}_{\rm c2}$(A$_1$) 
and $H^{\rm orb}_{\rm c2}$(A$_2$) for the A$_1$ and A$_2$ phases respectively.
Upper panel: The constructed phase diagram $H\parallel b$-axis in the $T$-$H$ plane.
$T_{\rm c2}$ starts at the negative $T$ and becomes positive at 10T, keeping up
until meeting the curve starting from $H^{\rm AUL}_{\rm c2}$. The A$_1$ phase with $T_{\rm c1}$=1.6K
disappears at $H$=23T.  Above $H_{\rm rot}$ the A$_2$ phase appears and vanishes at $H_{\rm m}$.
}
\end{figure}

We employ the same method to construct the phase diagram for $H\parallel b$-axis as in 
the main text except for $T_{\rm c2}<0$, thus we use $\kappa=3.9$K/$\mu_{\rm B}$.
As shown in Fig.~\ref{fig13}(b) which is identical with Fig.~\ref{fig1}(b), the magnetization curve $M_b(H)$ is plotted 
by the red line and the green line indicates $H_{\rm eff}$ for the A$_2$ phase.
The A$_1$ phase is essentially unchanged in the case of the main text,
namely it vanishes at $H$=22T corresponding to their orbital
limiting field $H^{\rm orb}_{\rm c2}(A_1)$ (see the gray band) 
because $H_{\rm eff}=H_{\rm ext}$ is unscreened.

In contrast, 
the A$_2$ phase starting with the negative temperature $T_{\rm c2}<0$
survives in much higher fields because $H_{\rm eff}$ denoted by the green line
comes back within the allowed region denoted by the blue band above the d-vector rotation
field $H>H_{\rm rot}$.
Thus it appears above $H_{\rm rot}$=15T first (see  in Fig.~\ref{fig13}(a))
and vanishes at the meta-magnetic transition $H_{\rm m}$=34T above which
as indicated by the green line it is outside the allowed region.

The main different from the case of $T_{\rm c2}>0$ lies in the fact that
until the $T_{\rm c2}$ becomes positive, the A$_2$ phase does not appear
and even when it becomes positive, it never appears because
$H_{\rm eff}(A_2)$ is outside the allowed region below $H_{\rm rot}$
(see the black straight line for $H=H_{\rm ext}$ in Fig.~\ref{fig13}(b)).

\begin{figure}
\begin{center}
\includegraphics[width=4cm]{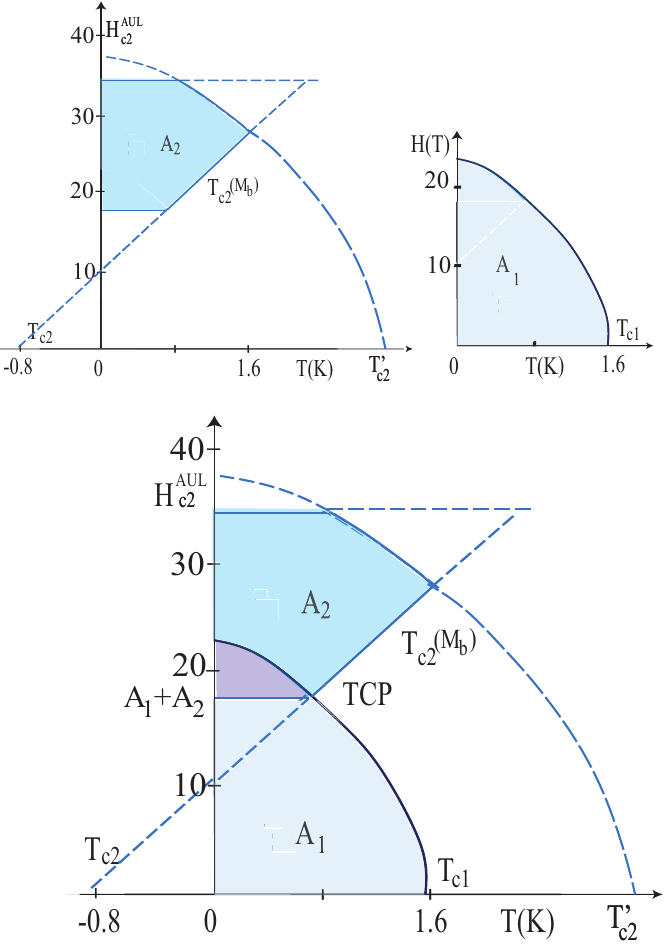}
\end{center}
\caption{\label{fig14}
%fig14(231113).pdf.
The upper panel shows the phase diagram for the A$_2$ phase (left) and the A$_1$ phase (right) in the $H$-$T$ plane.
The former does not contain the A$_2$ phase in the low $T$ and $H$ region. The lower panel shows the total phase diagram
combined with the two phases. The four second order lines meet at 15T, constituting the tetra-critical point.
The region around 20T consists of A$_1$ and A$_2$, giving the mixture phase.
}
\end{figure}
 
We show the A$_2$(left) and A$_1$(right) phases separately in the top panel of Fig.~\ref{fig14}.
It is clear that the A$_2$ phase is isolated and has a positive slope while the A$_1$ phase
has a simple conventional $H_{\rm c2}(T)$ form. The overlapping region of the two phases
gives rise to the mixture phase of A$_1$ and A$_2$ phases and
the tetra-critical point shown in the lower panel of Fig.~\ref{fig14}.
Note that in contrast with the main text with $T_{\rm c2}>0$ (see Fig.~\ref{fig2}), other internal transition line is absent
and the dotted lines are not realized.

\subsection{Phase diagram from $H\parallel b$-axis to $c$-axis}

\begin{figure}
\begin{center}
\includegraphics[width=6cm]{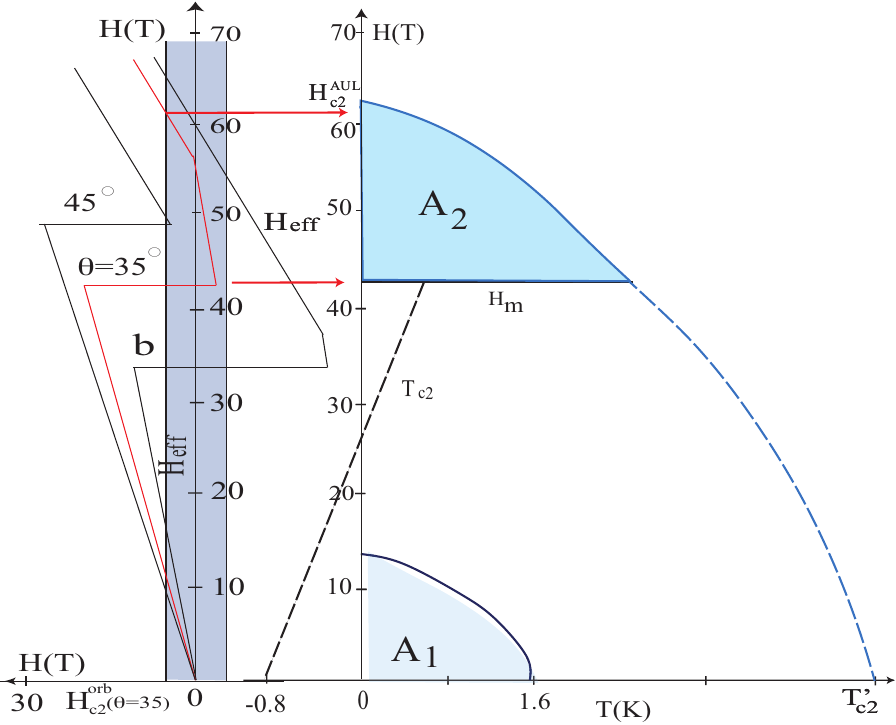}
\end{center}
\caption{\label{fig15}
%fig15(231113).pdf.
Phase diagram for $H\parallel \theta=35^{\circ}$. The left side panel shows $H_{\rm eff}$ by the red line with the
vertical allowed region bound by $H^{\rm orb}_{\rm c2}(\theta=35^{\circ})$.
The A$_2$ phase appears above $H_{\rm m}$ and below $H^{\rm AUL}_{\rm c2}$.
$T_{\rm c2}$ is not realized until it hits $H_{\rm m}$.
}
\end{figure}

We construct the phase diagram for the field direction tilted from the 
$b$-axis toward the $c$-axis by $\theta=35^{\circ}$.
As shown in the left panel of Fig.~\ref{fig15} by the red curve,
just above $H_{\rm m}$, $H_{\rm eff}$ becomes inside the allowed region denoted by the vertical band
which is defined by $\pm H^{\rm orb}_{\rm c2}(\theta=35^{\circ})$.
Thus the  A$_2$ phase, shown in the right hand side of Fig.~\ref{fig15},
appears above $H_{\rm m}$, a situation same as in the main text.
Here the straight dotted line starting at $T_{\rm c2}<0$ is never realized,
so there is no further internal transition line.

\section{Comparisons with ${\rm URhGe}$}

\begin{figure}
\begin{center}
\includegraphics[width=6cm]{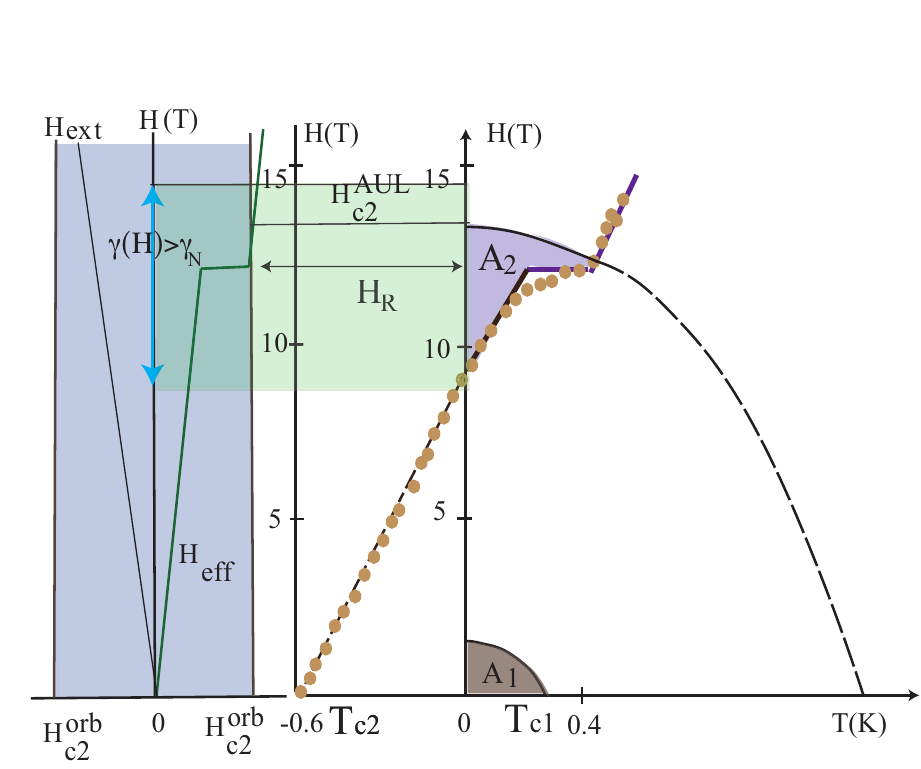}
\end{center}
\caption{\label{fig6}
%fig6(231109).pdf
Construction of the phase diagram for $H\parallel b$ for URhGe.
The left side panel: $H_{\rm eff}$ is shown in the green line relative to the
external field $H_{\rm ext}$ denoted by the black line, which runs out at $H^{\rm AUL}_{\rm c2}$
from the allowed region set by $H^{\rm orb}_{\rm c2}$.
The blue arrow indicates the region $\gamma(H)>\gamma_{\rm N}$, which limits the A$_2$ phase to appear.
The right side panel: $H$-$T$ phase diagram obtained. $T_{\rm c2}$=-0.6K starts from the negative value
according to $T_{\rm c2}=T_{\rm c0}+\kappa M_b(H)$, which becomes positive at 9T and jumps at the
moment rotation field $H_{\rm R}$. The A$_2$ phase appears around it, isolated from the A$_1$ phase
in low $H$ and $T$ region. This phase diagram is similar to that in UTe$_2$, having a portion ${dH_{\rm c2}\over dT}>0$.
}
\end{figure}

It is instructive to compare the above analysis on UTe$_2$ with URhGe~\cite{urhge0,urhge1,urhge2,machida3}
because they have similar phase diagrams in common under the field direction for the magnetic
hard $b$-axis.  
We construct the phase diagram for $H\parallel b$-axis in URhGe shown in Fig.~\ref{fig6}.
The construction procedure is the same as in the main text:
Since $T_{\rm c2}<0$ is situated deep in the negative side,
the $A_2$ phase appears when the magnetization $M_b$ exhibits a jump at $H_{\rm R}$
where the magnetic moment rotates from the magnetic easy $c$-axis.
$H_{\rm R}$ corresponds to the meta-magnetic transition $H_{\rm m}$ before.
We have chosen the following parameters to construct the phase diagrams here and throughout:
$T_{\rm c2}$=-0.6K under the observed transition temperature $T_{\rm c1}$=0.22K.
$\kappa$ is determined by the splitting of $T_{\rm c1}$-$T_{\rm c2}$=2$\kappa M_c$
with the spontaneous moment $M_c=0.4\mu_{\rm B}$, yielding $\kappa =1.0$K/$\mu_{\rm B}$.
Together with $\alpha^b_0=40.5$T/K, we use $H^{\rm orb}_{\rm c2}=\alpha^b_0|T_{\rm c2}|=24.3$T.

The magnetization jump greatly reduces the external field as depicted in the left side panel.
This results in that the isolated $A_2$ phase
appears above the $A_1$ phase stabilized only in lower fields.
$H^{\rm AUL}_{\rm c2}$ determined by  $H_{\rm eff}$ is comparable to the upper limit of the
extra DOS region ($\gamma(H)>\gamma_{\rm N}$) indicated by the blue
arrow along the $H$-axis of the left panel.
These two factors yield the upper bound in the $A_2$ phase.
It is rather remarkable to see that the magnetization curve of $M_b(H)$
indicated by the dots in the right panel forms the lower part of the reentrant 
$A_2$ phase. This proves that the expression $T_{\rm c2}=T_{\rm c0}+\kappa M_b$ is correct.

\begin{figure}
\begin{center}
\includegraphics[width=7cm]{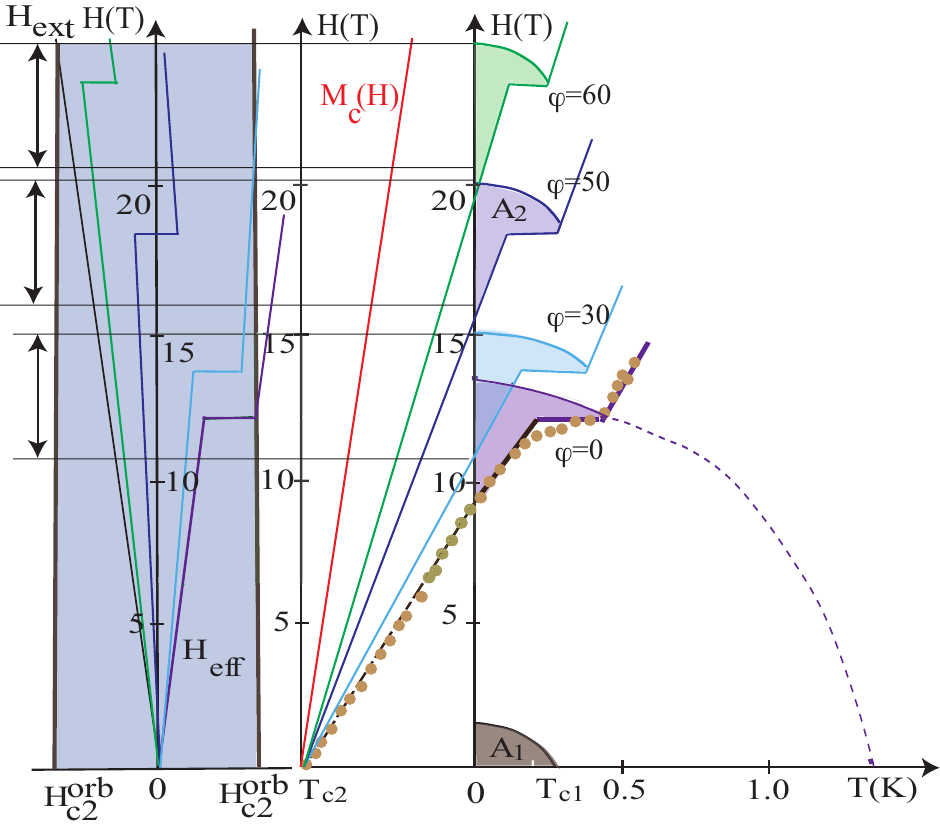}
\end{center}
\caption{\label{fig7}
%fig7(231108).pdf.
Construction of the phase diagrams under varying the angle $\varphi$ measured from
the $b$-axis toward the $a$-axis. The left side panel: $H_{\rm eff}$ for $\varphi=0^{\circ},
30^{\circ},50^{\circ}, 60^{\circ}$.
The three arrows indicate the field region for $\gamma(H)>\gamma_{\rm N}$ which limits the A$_2$ phase to appear.
The right side panel: $H$-$T$ phase diagram obtained where $T_{\rm c2}=T_{\rm c0}+\kappa M(\varphi)$ with $M(\varphi)$
depicted there for each $\varphi$ started from $T_{\rm c2}$. The reentrant A$_2$ phase persists up to higher angles, sticking with $H_{\rm R}$.
}
\end{figure}

By continuing the same arguments, we can construct the phase diagrams
when the field direction is tilted from the $b$-axis toward the other hard $a$-axis by the angle
$\varphi$. As displayed in the left hand panel of Fig.~\ref{fig7} where $H_{\rm eff}$ is shown for the
selected angles. To construct this figure, once the value $\kappa$ is fixed we need:\\
\noindent
(1) The magnetization curves of the $M_b(H)$ component for those angles $\varphi$ which are done by taking the
projection $M_b(H)$ onto the external field direction. $M_b(\varphi=\pi/2)$ coincides with $M_c(H)$.
This procedure is experimentally verified~\cite{miyake}.\\
\noindent
(2) $M_b(\varphi)$ starting from $T_{\rm c2}$ at the zero-field
yields $H_{\rm eff}=H+\alpha_0\kappa M_b(\varphi)$, which is shown in the right hand panel.
\noindent
(3) The initial slope $\alpha_0$ depends on the angle $\varphi$, but the $H_{\rm c2}$ anisotropy from
the $b$-axis to $a$-axis is weak, thus we neglect it. Note that this anisotropy was strong in UTe$_2$
which greatly influences the phase diagrams around $\theta\sim 40^{\circ}$ to attain the maximal $H_{\rm c2}(\theta)$
there.

It is now seen from Fig.~\ref{fig7}:\\
\noindent
(A) Since nearly all $H_{\rm eff}$ for various $\varphi$ are well within the allowed region up to the
highest field, meaning that $H^{\rm AUL}_{\rm c2}$ is not effective here. This is different from that in UTe$_2$.
 Instead, the upper limit of the $A_2$ phase is limited by the extra DOS with 
 $\gamma(H)>\gamma_{\rm N}$, whose field regions
are assumed to be constant here to be consistent with the experimental data.
Namely the width of the field region for $\gamma(H)>\gamma_{\rm N}$ is a constant centered around $H_{\rm R}(\varphi)$.
This assumption should be checked experimentally. Indirect supporting evidence comes from
the pressure experiment by Miyake et al~\cite{urhge2}, showing that the shape of the reentrant $A_2$ phases
under pressure simply shifts without changing their width along the field direction, which is consistent with
our assumption. \\
\noindent
(B) The magnetization jump always occurs when the magnetization $M_{\varphi}(H)$  reaches at a certain value
$\sim 0.4\mu_{\rm B}$ for all $\varphi$. This is 
consistent with the experimental observation $H_{\rm R}(\theta)\propto 1/\cos\varphi$.
This means that in the right hand panel in Fig.~\ref{fig7} the jump point for each $\varphi$
aligns vertically parallel to the $H$-axis. This guarantees the existence of the reentrant $A_2$ phase to always 
appear just around their $H_{\rm R}$. This is the reason why it is accompanied by $H_{\rm R}$ up to a higher angle $\varphi$.

\begin{figure}
\begin{center}
\includegraphics[width=5cm]{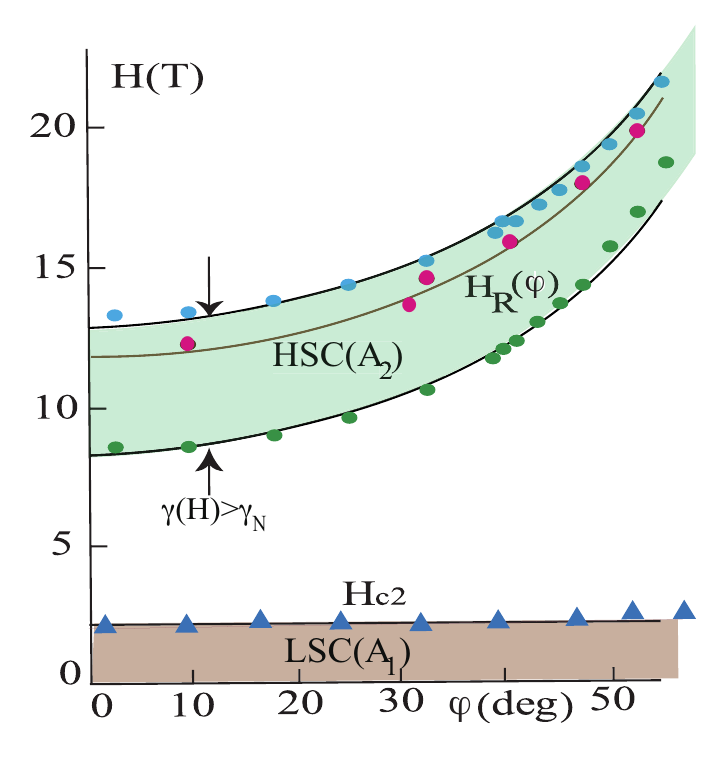}
\end{center}
\caption{\label{fig8}
%fig8(231108).pdf.
The constructed phase diagram in $H$-$\varphi$ plane.
The reentrant HSC(A$_2$) is within the light green band around $H_{\rm rot}(\varphi)$ denoted by a thin black curve
with red dots. The blue, and green dots and triangle symbols are experimental data for superconducting transitions.
The two arrows indicate the width of the field region for $\gamma(H)>\gamma_{\rm N}$.
LSC(A$_1$) is almost isotropic against $\varphi$.
}
\end{figure}

As shown in Fig.~\ref{fig8}, the $A_2$ phase forms a band centered around $H_{\rm R}$
at least up to $\varphi=50^{\circ}$. This situation is quite different from that in UTe$_2$
shown in Fig.~\ref{fig5}. In short, in the latter case the two factors $H^{\rm AUL}_{\rm c2}$
and the extra DOS region compete with each other to confine the $A_2$ phase to a narrow angle region.
However the attained field of $H_{\rm c2}$ is far high because the magnetization jump is bigger
0.5$\mu_{\rm B}$ in UTe$_2$ than $\sim 0.1\mu_{\rm B}$ in URhGe.

\section{Multiple phase diagrams under pressure}

We have postulated the existence of the second transition $T_{\rm c2}$=0.4K at $H$=0 to construct the
phase diagram for the $H\parallel b$-axis. It implies that the $H$-$T$ phase diagrams for all field directions
should consist of multiple phases in general. Here we demonstrate that this assumption is quite natural
when analyzing the phase diagrams under pressure. Moreover, the third phase $A_0$, which is hidden 
so far appears to exist, completing the three component scenario based on SO(3)$^{\rm spin}$ introduced previously.

\begin{figure}
\begin{center}
\includegraphics[width=7cm]{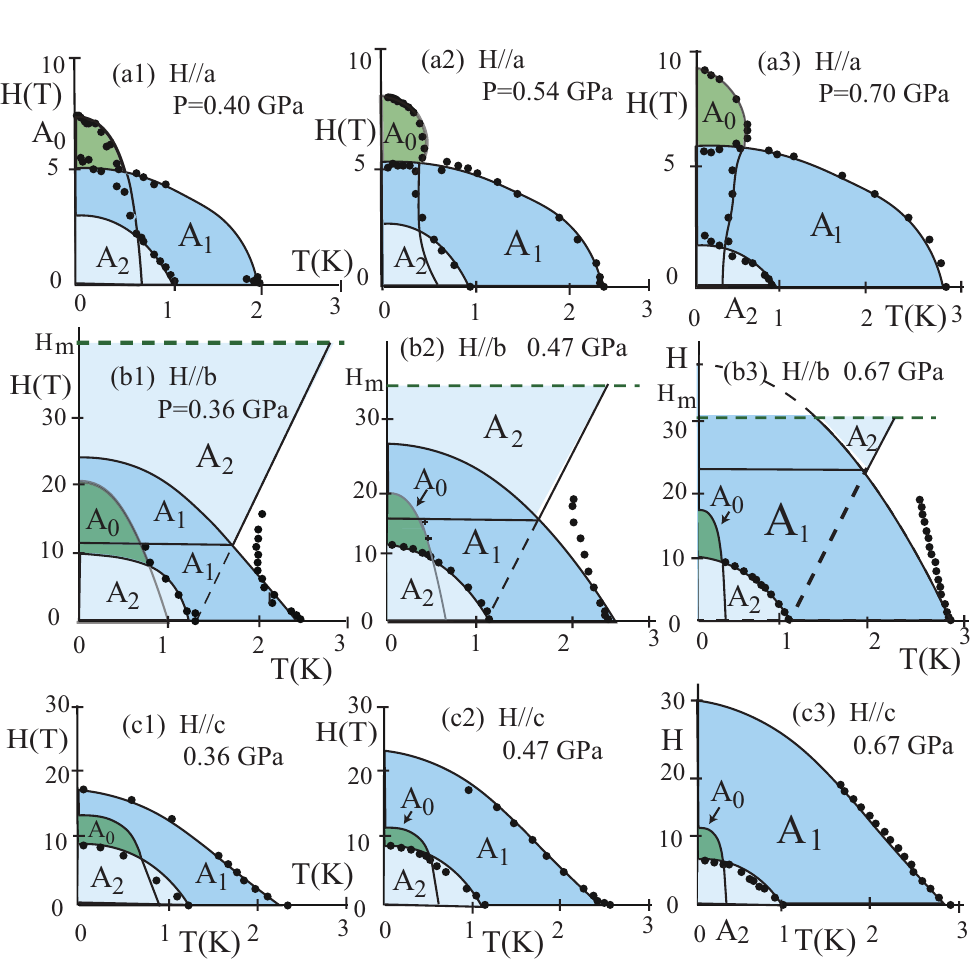}
\end{center}
\caption{\label{fig9}
%fig9(231119).pdf. 
Phase diagrams under pressure~\cite{pressure1,pressure2,pressure3,pressure4,butch1,butch2,butch3}. 
Each row corresponds to the field orientations $a$-axis, $b$-axis, and $c$-axis.
Each column corresponds to approximately the same pressure, low, middle, and high pressures. The black dots are experimental data points. The identified phases for A$_1$, A$_2$, and A$_0$ are colored differently.
The produced phase diagrams capture the experimental data points. Systematic evolutions of the three phases are
seen under pressures and field orientations.
}
\end{figure}

In Fig.~\ref{fig9}, we analyze the $H$-$T$ phase diagrams for all principal field directions; $a$, $b$ and $c$-axes
under pressure~\cite{pressure1,pressure2,pressure3,pressure4,butch1,butch2,butch3}. It is seen from Fig.~\ref{fig9} that:\\
\noindent
(1) It is obvious that all phase diagrams consist of the multiple phases,
independent of $P$ and field orientations.\\
\noindent
(2) The multiple phases are classified by the three phases, $A_1$, $A_2$, and $A_0$.
There is no further phase existed in this analysis. This is consistent with our three component
scenario.\\
\noindent
(3) The $A_0$ phase is somewhat hidden because so far at $H=0$ there is no direct experimental report 
to third transition $T_{\rm c3}$ while the double transition $T_{\rm c1}$ and $T_{\rm c2}$ are confirmed
by several experiments mentioned above. However, if we look carefully at the data for $H\parallel a$-axis 
in Figs.~\ref{fig9} (a1), (a2), and (a3),
 it is understood that 
 there must be an additional phase in high fields, namely the $A_0$ phase indicated by green,
 which is systematically evolving under $P$.\\
\noindent
(4) As for the field direction $b$-axis shown in Figs.~\ref{fig9} (b1), (b2), and (b3),
we see that the $A_2$ phase starting at $T_{\rm c2}$  quickly disappears at lower fields.
In contrast, the high temperature $A_1$ phase starting at $T_{\rm c1}$
exhibits a strong rise, and a positive slope in the middle field region
toward $H_{\rm m}$ indicated by horizontal dotted lines.
The other dotted lines there from $T_{\rm c2}$ toward the tetra-critical points are not realized.
The existing data points are completely exhausted by those analyses, so 
we understand that there are no positive sloped internal phase transition lines inside the $A_1$ phase.\\
\noindent
(5) This assignment is completely consistent with the constructed  $b$-axis phase diagram
shown in Figs.~\ref{fig1}, ~\ref{fig2}, ~\ref{fig3}, and~\ref{fig4} in the ambient pressure.
Therefore, as a function of $P$, including the ambient pressure the phase diagrams are
continuously evolving.\\
\noindent
(6) We notice that the small pocket regions for the A$_2$ phase 
around the origin always exist, independent of $P$ and field directions,
a situation quite similar to the ambient pressure as shown in Figs.~\ref{fig2}(c) and (d).

To conclude this section, through the examinations of the data under pressure,
it is natural to assume that in the ambient pressure $T_{\rm c2}$
exists at $H$=0 for $T_{\rm c}$=2.1K samples.  The $A_2$ phase 
at low $T$ and $H$ occupied a small pocket area in the $H$-$T$ phase diagram shown in
Figs.~\ref{fig1}$\sim$\ref{fig4}.
This is in contrast with $T_{\rm c}$=1.6K samples where
$T_{\rm c2}$ is absent, or negative because the large residual DOS wipes out the
$A_2$ phase to appear.

\section{Available DOS for $A_2$ phase in high fields
--A hint for pairing mechanism--}

\begin{figure}
\begin{center}
\includegraphics[width=6cm]{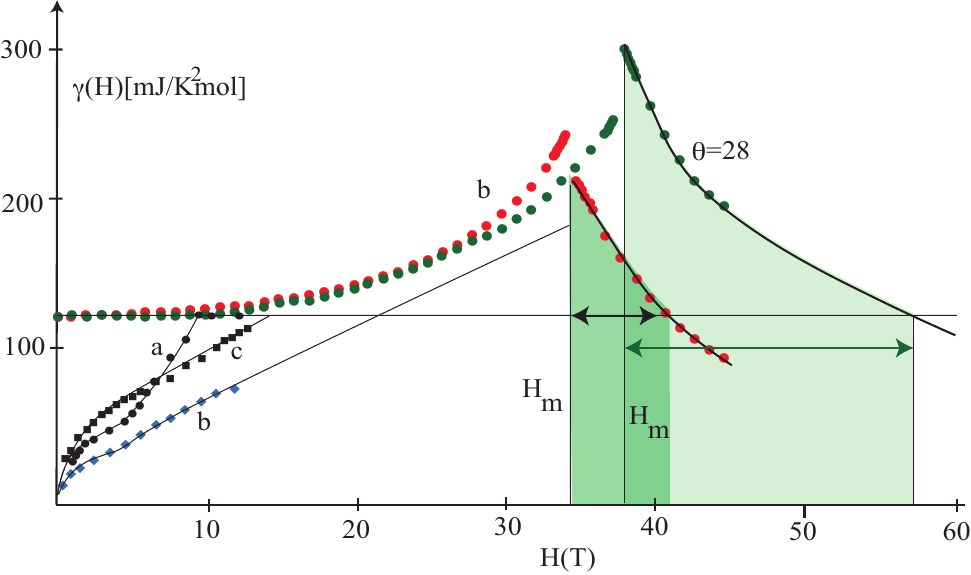}
\end{center}
\caption{\label{fig10}
%fig10(231119).pdf.
Field dependences~\cite{miyake2} of the normal DOS $\gamma(H)$ for $H\parallel b$-axis and tilted by $\theta=28^{\circ}$.
Both show the enhancements toward $H_{\rm m}$ and decreases. 
The colored regions ($H>H_{\rm m}$) with the arrows
show the enhancement
exceeded above $\gamma_{\rm N}=120{\rm mJ}/{\rm K}^2 {\rm mol}$ and 
differ for the two cases where the latter is far wider than the former.
The low field data below $\gamma_{\rm N}$ show the superconducting $\gamma(H)$ for three orientations~\cite{roman}.
}
\end{figure}

Here in Fig.~\ref{fig10} we refer to the experimental data for the normal DOS for the $b$-axis and
$\theta=28^{\circ}$ by Miyake et al~\cite{miyake2}. The DOS or $\gamma(H)$ enhancement occurs 
associated with the meta-magnetic transition at $H_{\rm m}$.
Above $H_{\rm m}$, as a function of $H$, $\gamma(H)$ quickly decreases.
Comparing the field regions, in which $\gamma(H)>\gamma_{\rm N}$ indicated by
the horizontal arrows are far wider for the $\theta=28^{\circ}$case than that in the $b$-axis case.

On the other hand, $\gamma(H)$'s in the superconducting state for the three $a$, $b$, and 
$c$ axes~\cite{roman} are also shown in Fig.~\ref{fig10} by the black dots. 
When $\gamma(H)$ reaches $\gamma_{\rm N}$=120mJ/molK$^2$
for $a$ and $c$-axes, the system becomes normal. Extrapolating the $b$-axis data linearly to higher 
fields, it reaches $\gamma_{\rm N}$ around 22T or so at which the $A_1$ phase disappears as
seen from Fig.~\ref{fig2} simply because the $A_1$ phase exhausts its available DOS, e.g. $\gamma_{\rm N}$. 
Thus for the $A_2$ phase to reappear around that field region,
it is necessary to be available for the extra DOS, otherwise, there is no usable DOS for the $A_2$ phase
or no room for the $A_2$ phase.
Then as mentioned above according to the DOS enhancement associated with the
meta-magnetic transition, the extra DOS becomes available for the $A_2$ phase
around $H_{\rm m}$. This field interval is necessary for the $A_2$ phase to reappear in
high fields for $\theta\sim35^{\circ}$ as explained in the previous section.
Note that for $H\parallel b$-axis this field interval is too narrow for the $A_2$ phase to reappear
above $H_{\rm m}$ even though $H^{\rm AUL}_{\rm c2}(\rm lower)<H_{\rm eff}<H^{\rm AUL}_{\rm c2}(\rm upper)$. 
The enhanced DOS is quite high for $\theta=28^{\circ}$ 
which is more than $2\times\gamma_{\rm N}$,
making the extrapolated $T_{\rm c2}'\sim6$K so high as shown in Figs.~\ref{fig3} and ~\ref{fig4}.
It is also noticed that in URhGe high field reentrant $A_2$ phase is mainly constrained
by this extra DOS interval.
This suggests that the pairing mechanism is closely related to the origin of the mass enhancement.
The reentrant A$_2$ phase in URhGe is always associated with this mass enhancement region.

\section{Estimate for the KS value for the $a$-axis and the Pauli paramagnetic field}

The new KS value for the $a$-axis~\cite{matsumura}
 at lower $H$ is $\Delta K_a$=4.8$\%$
where the spin part of the KS $K_{\rm spin}$=1.3$\%$, namely,
$\Delta K_a/K_{\rm spin}$=3.7.
This evaluation is consistent with the following estimate:
The induced magnetization at $H$=1T is $M_a=-4.8\%\times 1{\rm T}/3.8{\rm T}/\mu_{\rm B}
=-0.00126\mu_{\rm B}$
with the hyperfine coupling constant $A^{\rm hyperfine}_a=3.8{\rm T}/\mu_{\rm B}$ for the Te site 2.
On the other hand, the Pauli paramagnetic moment $M_{\rm Pauli}$ at $H$=1T
is given by $M_{\rm Pauli}=\chi_{\rm Pauli}\times 1{\rm T}=2\mu^2_{\rm B}N(0)\times 1{\rm T}=0.003\mu_{\rm B}$.
Thus, $M_a/M_{\rm Pauli}=-0.0126\mu_{\rm B}/0.003\mu_{\rm B}$=-4.2
where $N(0)=120{\rm mJ}/{\rm molK}^2$.
Both estimates yield the Knight shift value -3.7$\sim$-4.2$\chi_{\rm N}$.
Note that contrary to this anomalously large KS, for the $b$ and $c$-axes
the corresponding values are the following:
$\Delta K_b/K_{\rm spin}\sim\Delta K_c/K_{\rm spin}$=-0.15
and $M_b/M_{\rm Pauli}$=-0.13 and $M_c/M_{\rm Pauli}$=-0.17.

From the observed KS value we can evaluate the Pauli paramagnetic limiting field
as $H^a_{\rm Pauli}$=2T, which is compared with $H^b_{\rm Pauli}$=10T.
These values are far below the observed $H_{\rm c2}$ for the $a$ and $b$-axes.
This strengthens the scenario based on a triplet pairing for UTe$_2$.

\section{Different estimate for the KS value for the $a$-axis }

According to Miyake~\cite{miyakeriron}, the induced magnetization value associated with the
A$_1$ phase is evaluated by a formula relative to the normal state due to the
Pauli paramagnetism $M_{\rm Pauli}$:

\begin{eqnarray}
{|\delta M| \over M_{\rm Pauli}}={\alpha \Delta^2\over 4\epsilon_{\rm F}\mu_{\rm B}H}(1+{2\over gN(0)})
(1+F^a_0)
\label{chi}
\end{eqnarray}

\noindent
where $\alpha\sim O(1)$ is the parameter to express the degree of the electron-hole asymmetry,
namely $N'(0)$ the derivative at the Fermi level $\epsilon_{\rm F}(=T_{\rm K}=30$K). $gN(0)$ is the dimensionless 
coupling constant, assuming here 0.14. $F^a_0$ is the Landau Fermi liquid parameter, which we ignore.
For $\Delta =1.75T_{\rm c1}$(K), and $H$=1T, we find $|\delta M|/ M_{\rm Pauli}=2.2\alpha$.
This is the same order of magnitude estimated in the main text.

 \section{Evaluation of $\gamma(H)$ from the jump of the specific heat at $T_{\rm c}$}

\begin{figure}
\begin{center}
\includegraphics[width=4cm]{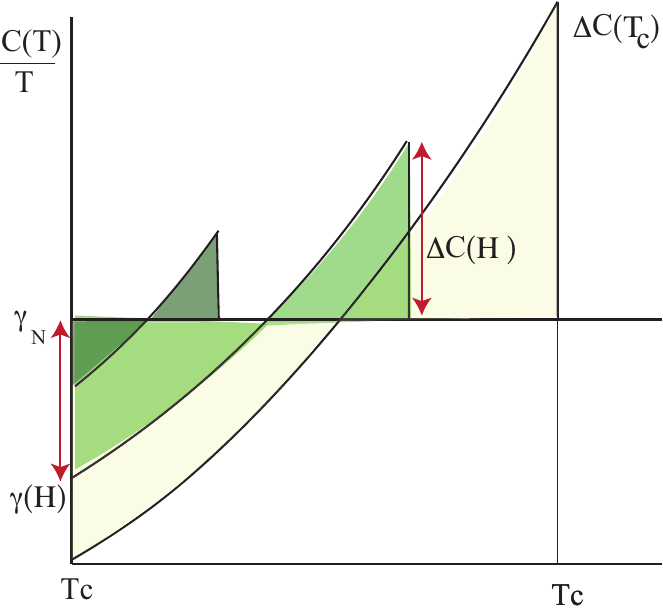}
\end{center}
\caption{\label{fig16}
%fig16(231113).pdf: 
The specific heat jump $\Delta C(H)$ at $T_{\rm c}(H)$ under fields 
indicated by the red arrow. The entropy balance requires that
the two triangle areas above and below $\gamma_{\rm N}$ must be equal,
meaning that the length $\gamma_{\rm N} - \gamma(H)$ is uniquely 
related to $\Delta C(H)$, independent of the functional form of $C(H)$.
}
\end{figure}

 We can estimate $\gamma(H)$ from the specific heat jump $\Delta C(H)$ at $T_{\rm c}$ by assuming the
 entropy balance between the SC and the normal state as follows:
 As shown in Fig.~\ref{fig16},
 at $H=0$, $\gamma(H=0)=0$ and $\Delta C(H=0)=1.43\gamma_{\rm N}$ for the BCS.
 As $H$ increase from zero, the jump decrease, and the residual $\gamma(H)$ increases
 so that both keep the entropy balance, namely the area of the upper triangle equals that of the lower triangle.
 Then we can derive a formula under the assumption that all the specific heat curves are scaled:
 
\begin{eqnarray}
{\Delta C(H) \over \Delta C_{\rm BCS}}=1-{\gamma(H)\over \gamma_{\rm N}}.
\label{chi}
\end{eqnarray}
 
 \noindent
 Since this relation is independent of the functional form of the $C(T)$ curve 
 and the magnitude of the jump, 
 we can generalize it to the non-BCS cases as
  
\begin{eqnarray}
{\gamma(H)\over \gamma_{\rm N}}=1-{\Delta C(H) \over \Delta C(H=0)}.
\label{chi}
\end{eqnarray}
 
\noindent
In fact we confirm this formula for UTe$_2$ by comparing the estimates ${\gamma(H)/\gamma_{\rm N}}$ 
calculated by this formula with the measured values~\cite{roman} as displayed in Fig.~\ref{fig12} where the black arrows
indicates the jumps $\Delta C(T_{\rm c})$ in the $H\parallel b$-axis~\cite{rosuel}.
It is seen that they agree with each other rather perfectly.
This tells us that the analysis of the specific heat jump associated with the A$_2$ phase in the double transition
around 17T gives reliable estimates for the $\gamma(A_2)$ which are also displayed in Fig.~\ref{fig12}
as the red arrows.


\begin{thebibliography}{9}

\bibitem{ran}
Sheng Ran, Chris Eckberg, Qing-Ping Ding, Yuji Furukawa, Tristin Metz, Shanta R. Saha, I-Lin Liu, Mark Zic, Hyunsoo Kim, Johnpierre Paglione, Nicholas P. Butch,
Nearly ferromagnetic spin-triplet superconductivity,
Science {\bf 365}, 684 (2019).

\bibitem{aoki0}
D. Aoki, A. Nakamura, F. Honda, D. Li, Y. Homma, Y. Shimizu, Y. J. Sato, G. Knebel, J.-P. Brison, A. Pourret, D. Braithwaite, G. Lapertot, Q. Niu, M. Vali\v{s}ka, H. Harima, and J. Flouquet, 
Unconventional Superconductivity in Heavy Fermion UTe$_2$,
J. Phys. Soc. Jpn. {\bf 88}, 043702 (2019).

\bibitem{review}
D. Aoki, J. -P. Brison, J. Flouquet, K. Ishida, G. Knebel, Y. Tokunaga, and Y. Yanase,
Unconventional Superconductivity in UTe$_2$,
J. Phys.: Condens. Matter {\bf 34}, 243002 (2022).

\bibitem{review1}
S.K. Lewin,  C.E. Frank,  S. Ran, J. Paglione,  and N.P. Butch,  
A Review of UTe$_2$ at High Magnetic Fields, 
Rep. Prog. Phys. {\bf 86}, 114501 (2023).
 

\bibitem{taillefer}
R. Joynt and L. Taillefer,
The superconducting phases of UPt$_3$,
Rev. Mod. Phys. {\bf 74}, 235 (2002).

\bibitem{sauls}
J. A. Sauls,
The Order Parameter for the Superconducting Phases of UPt$_3$,
Adv. Phys. {\bf 43}, 113 (1994).


\bibitem{upt3}
K. Machida and M. Ozaki,
Superconducting double transition in a heavy-fermion material UPt$_3$,
Phys. Rev. Lett. 66, 3293 (1991).

\bibitem{ohmi}
T. Ohmi and K. Machida,
Nonunitary superconducting state in UPt$_3$,
Phys. Rev. Lett. 71, 625 (1993).

\bibitem{yo}
Y. Machida, A. Itoh, Y. So, K. Izawa, Y. Haga, E. Yamamoto, 
N. Kimura, Y. Onuki, Y. Tsutsumi, and K. Machida, 
Twofold Spontaneous Symmetry Breaking in the Heavy-Fermion Superconductor UPt$_3$,
Phys. Rev. Lett. {\bf 108},  175002 (2012).

\bibitem{tsutsumi1}
Y. Tsutsumi, M. Ishikawa, T. Kawakami, T. Mizushima, M. Sato, M. Ichioka, and K. Machida,
UPt$_3$ as a Topological Crystalline Superconductor,
J. Phys. Soc. Jpn. {\bf 82}, 113707 (2013).

\bibitem{ott}
H. R. Ott, H. Rudigier, T. M. Rice, K. Ueda, Z. Fisk, and J. L. Smith,
p-Wave Superconductivity in UBe$_{13}$,
Phys. Rev. Lett. {\bf 52}, 1915 (1984).

\bibitem{shimizu1}
Yusei Shimizu, Daniel Braithwaite, Dai Aoki, Bernard Salce, and Jean-Pascal Brison,
Spin-Triplet p-Wave Superconductivity Revealed under High Pressure in 
UBe$_{13}$,
Phys. Rev. Lett. {\bf 122}, 067001 (2019).

\bibitem{shimizu2}
Yusei Shimizu, Shunichiro Kittaka, Toshiro Sakakibara, Yoshinori Haga, Etsuji Yamamoto, Hiroshi Amitsuka, Yasumasa Tsutsumi, and Kazushige Machida,
Field-Orientation Dependence of Low-Energy Quasiparticle Excitations in the Heavy-Electron Superconductor 
UBe$_{13}$,
Phys. Rev. Lett. {\bf 114}, 147002 (2015).

\bibitem{Th1}
J. E. Sonier, R. H. Heffner, D. E. MacLaughlin, G. J. Nieuwenhuys, O. Bernal, 
R. Movshovich, P. G. Pagliuso, J. Cooley, J. L. Smith, and J. D. Thompson,
$\mu^+$ Knight Shift Measurements in U$_{0.965}$Th$_{0.035}$Be$_{13}$ Single Crystals,
Phys. Rev. Lett. {\bf 85}, 2821 (2000).

\bibitem{shimizu3}
Yusei Shimizu, Shunichiro Kittaka, Shota Nakamura, Toshiro Sakakibara, Dai Aoki, Yoshiya Homma, Ai Nakamura, and Kazushige Machida,
Quasiparticle excitations and evidence for superconducting double transitions in monocrystalline U$_{0.97}$Th$_{0.03}$Be$_{13}$,
Phys. Rev. B {\bf 96}, 100505 (2017).

\bibitem{Th}
Kazushige Machida,
Spin Triplet Nematic Pairing Symmetry and Superconducting Double Transition in U$_{1-x}$Th$_{x}$Be$_{13}$,
J. Phys. Soc. Jpn. {\bf 87}, 033703 (2018).


\bibitem{3he}
D. Vollhart and P. W{\"o}lfle,
The superfluid phases of Helium 3,
Taylor and Francis, London, 1990.


\bibitem{af1}
C. Duan, K. Sasmal, M. B. Maple,  A. Podlesnyak,  J.-X. Zhu,  Q. Si, and P.  Dai, 
Incommensurate Spin Fluctuations in the Spin-triplet Superconductor Candidate UTe$_2$, 
Phys. Rev. Lett.  {\bf 125}, 237003 (2020).

\bibitem{af2}
W. Knafo,G. Knebel, P. Steffens, K. Kaneko, A. Rosuel, J.-P. Brison, J. Flouquet, D. Aoki,  G. Lapertot, and S. Raymond, 
Low-dimensional antiferromagnetic fluctuations in the heavy-fermion paramagnetic ladder UTe$_2$, 
Phys. Rev. B {\bf 104}, L100409 (2021).

\bibitem{af3}
N. P. Butch,  S. Ran,  S. R. Saha,  P. M. Neves, M. P. Zic,  J. Paglione,  S. Gladchenko,   Q. Ye, and J. A. Rodriguez-Rivera,  Symmetry of magnetic correlations in spin-triplet superconductor UTe$_2$,
 npj Quantum Mater. {\bf 7}, 39 (2022).


\bibitem{sakai}
H. Sakai, Y. Tokiwa, P. Opletal, M. Kimata, S. Awaji, T. Sasaki, D. Aoki, S. Kambe, Y. Tokunaga, and Y. Haga,
Field Induced Multiple Superconducting Phases in UTe$_2$ along Hard Magnetic Axis,
Phys. Rev. Lett. {\bf 130}, 196002 (2023).


\bibitem{sakai2.1K}
H. Sakai, P. Opletal, Y. Tokiwa, E. Yamamoto, Y. Tokunaga, S. Kambe, and Y. Haga, 
Single crystal growth of superconducting 
UTe$_2$ by the molten salt flux method,
Phys. Rev. Mater. {\bf 6}, 073401 (2022).


\bibitem{rosuel}
A. Rosuel, C. Marcenat, G. Knebel, T. Klein, A. Pourret, N. Marquardt,
Q. Niu, S. Rousseau, A. Demuer, G. Seyfarth, G. Lapertot, D. Aoki, D. Braithwaite, J. Flouquet, and
 J. -P. Brison, 
 Field-induced tuning of the pairing state in a superconductor,
Phys. Rev. X {\bf 13}, 011022 (2023).


\bibitem{machida1}
K. Machida,
Theory of Spin-polarized Superconductors--An Analogue of Superfluid $^3$He A-phase--,
J. Phys. Soc. Jpn. {\bf 89}, 033702 (2020).

\bibitem{machida2}
K. Machida,
Notes on Multiple Superconducting Phases in UTe$_2$ --Third Transition--,
J. Phys. Soc. Jpn. {\bf 89}, 0655001 (2020).


\bibitem{machida3}
K. Machida,
Nonunitary triplet superconductivity tuned by field-controlled magnetization: 
URhGe, UCoGe, and UTe$_2$,
Phys. Rev. B {\bf 104}, 014514 (2021).

\bibitem{machida4}
K. Machida,
Violation of Pauli-Clogston limit in the heavy-fermion superconductor CeRh$_2$As$_2$: 
Duality of itinerant and localized 4f electrons,
Phys. Rev. B {\bf 106}, 184509 (2022).



\bibitem{machida5}
K. Machida,
Violation of the orbital depairing limit in a nonunitary state:
High-field phase in the heavy fermion superconductor UTe$_2$,
Phys. Rev. B {\bf 107}, 224512 (2023).


\bibitem{lewin}
Sylvia K. Lewin, Peter Czajka, Corey E. Frank, Gicela Saucedo Salas, Hyeok Yoon, Yun Suk Eo, Johnpierre Paglione, Andriy H. Nevidomskyy, John Singleton, and Nicholas P. Butch,
High-Field Superconducting Halo in UTe$_2$,
arXiv:2402.18564.



\bibitem{matsumura}
Hiroki Matsumura, Hiroki Fujibayash, Katsuki Kinjo, Shunsaku Kitagawa, Kenji Ishida, 
Yo Tokunaga, Hironori Sakai, Shinsaku Kambe, Ai Nakamura, Yusei Shimizu, Yoshiya Homma, 
Dexin Li, Fuminori Honda, and Dai Aoki,
Large Reduction in the a-axis Knight Shift on UTe$_2$ with $T_{\rm c}$ = 2.1 K,
J. Phys. Soc. Japan {\bf 92},  063701 (2023).

\bibitem{ishida1}
G. Nakamine, Shunsaku Kitagawa, Kenji Ishida, Yo Tokunaga, 
Hironori Sakai, Shinsaku Kambe, Ai Nakamura, Yusei Shimizu, Yoshiya Homma, Dexin Li, Fuminori Honda, and Dai Aoki,
Superconducting properties of heavy fermion UTe$_2$ revealed by $^{125}$Te-nuclear magnetic resonance,
J. Phys. Soc. Jpn. {\bf 88}, 113703 (2019).

\bibitem{ishida2}
Genki Nakamine, Katsuki Kinjo, Shunsaku Kitagawa, Kenji Ishida, Yo Tokunaga, Hironori Sakai, 
Shinsaku Kambe, Ai Nakamura, Yusei Shimizu, Yoshiya Homma, Dexin Li, Fuminori Honda, and Dai Aoki,
Inhomogeneous Superconducting State Probed by $^{125}$Te NMR on UTe$_2$,
J. Phys. Soc. Japan {\bf 90},  064709 (2021).

\bibitem{ishida3}
Genki Nakamine, Katsuki Kinjo, Shunsaku Kitagawa, Kenji Ishida, Yo Tokunaga, Hironori Sakai, Shinsaku Kambe, Ai Nakamura, Yusei Shimizu, Yoshiya Homma, Dexin Li,  Fuminori Honda, and Dai Aoki,
Anisotropic response of spin susceptibility in the superconducting state of UTe$_2$
 probed with $^{125}$Te-NMR measurement,
 Phys. Rev. B {\bf 103}, L100503 (2021).
 
 \bibitem{ishida4}
 Hiroki Fujibayashi, Genki Nakamine, Katsuki Kinjo, Shunsaku Kitagawa, Kenji Ishida1, Yo Tokunaga, 
 Hironori Sakai, Shinsaku Kambe, Ai Nakamura, Yusei Shimizu, Yoshiya Homma, Dexin Li, Fuminori Honda, and Dai Aoki,
 Superconducting Order Parameter in UTe$_2$ Determined by Knight Shift Measurement,
  J. Phys. Soc. Jpn. {\bf 91}, 043705 (2022).
  
 \bibitem{ishida5}
K. Kinjo, H. Fujibayashi, S. Kitagawa, K. Ishida, Y. Tokunaga,  H. Sakai, S. Kambe, A. Nakamura,
Y. Shimizu, Y. Homma, D. X. Li, F. Honda, D. Aoki,  K. Hiraki,  M. Kimata,  and T. Sasaki,
Change of superconducting character in UTe$_2$ induced by a magnetic field, 
 Phys. Rev. B {\bf 107}, L060502 (2023).


 
\bibitem{kinjo}
Katsuki Kinjo, Hiroki Fujibayashi, Hiroki Matsumura, Fumiya Hori, Shunsaku Kitagawa,
Kenji Ishida, Yo Tokunaga, Hironori Sakai, Shinsaku Kambe, Ai Nakamura, Yusei Shimizu,
Yoshiya Homma, Dexin Li, Fuminori Honda, and Dai Aoki,
Superconducting spin reorientation in spin-triplet
multiple superconducting phases of UTe$_2$,
Sci. Adv. {\bf 9}, eadg 2736 (2023).

\bibitem{ran2}
S. Ran, I. L. Liu, Y. S. Eo, D. J. Campbell, P. M. Neves, W. T. Fuhrman, S. R. Saha, 
C. Eckberg, H. Kim, D. Graf, F. Balakirev, J. Singleton, J. Paglione, and N. P. Butch, 
Extreme magnetic field-boosted superconductivity, 
Nat. Phys. {\bf 15}, 1250 (2019).

\bibitem{georg}
G. Knebel, W. Knafo, A. Pourret, Q. Niu, M. Vali\v{s}ka, D. Braithwaite, 
G. Lapertot, M. Nardone, A.  Zitouni, S. Mishra, I. Sheikin, G. Seyfarth, J.-P. Brison, D. Aoki, and J. Flouquet,
Field-reentrant superconductivity close to a metamagnetic transition in the
heavy-fermion superconductor UTe$_2$, 
J. Phys. Soc. Jpn. {\bf 88}, 063707 (2019).

 
\bibitem{helm}
T. Helm, M. Kimata, K. Sudo, A. Miyata, J. Stirnat, T. F\"orster, J. Hornung, M. K\"onig, 
I. Sheikin, A. Pourret, G.
Lapertot, D. Aoki, J. -P. Brison, G. Knebel, and J. Wosnitza,
Suppressed magnetic scattering sets conditions for the emergence of 40T high-field 
superconductivity in UTe$_2$,
arXiv:2207.08261.


\bibitem{lonzarich}
Z. Wu, T. I. Weinberger, J. Chen, A. Cabala, D. V. Chichinadze, D. Shaffer, J. Pospisil, 
J. Prokleska, T. Haidamak, G. Bastien, V. Sechovsky, A. J. Hickey, 
M. J. Mancera-Ugarte, S. Benjamin, D. E. Graf, Y. Skourski, G. G. Lonzarich, M. Valiska, F. M. Grosche, and A. G. Eaton,
Enhanced triplet superconductivity in next-generation ultraclean UTe$_2$,
arXiv:2305.19033.


\bibitem{urhge0}
D. Aoki, K. Ishida, and J. Flouquet, 
Review of U-based Ferromagnetic Superconductors:
Comparison between UGe$_2$, URhGe, and UCoGe,
J. Phys. Soc. Jpn. {\bf 88}, 022001 (2019).


\bibitem{urhge1}
A. Miyake, D. Aoki, and J. Flouquet, Pressure evolution of
the ferromagnetic and field re-entrant superconductivity in
URhGe, J. Phys. Soc. Jpn. {\bf 78}, 063703 (2009).

\bibitem{urhge2}
D. Braithwaite, D. Aoki, J.-P. Brison, J. Flouquet, G. Knebel, A. Nakamura, and A. Pourret, 
Dimensionality Driven Enhancement of Ferromagnetic Superconductivity in URhGe,
Phys. Rev. Lett. {\bf 120}, 037001 (2018).



\bibitem{mizushima}
Takeshi Mizushima, Yasumasa Tsutsumi, Takuto Kawakami, Masatoshi Sato, Masanori Ichioka, and Kazushige Machida,
Symmetry-Protected Topological Superfluids and Superconductors ---From the Basics to $^3$He---,
J. Phys. Soc. Jpn. {\bf 85}, 022001 (2016).



\bibitem{pressure1}
 Dai Aoki, Fuminori Honda, Georg Knebel, Daniel Braithwaite, Ai Nakamura, DeXin Li, Yoshiya Homma, 
 Yusei Shimizu, Yoshiki J. Sato, Jean-Pascal Brison, and Jacques Flouquet,
 Multiple Superconducting Phases and Unusual Enhancement of the Upper Critical Field in UTe$_2$,
 J. Phys. Soc. Jpn. {\bf 89}, 053705 (2020).

\bibitem{pressure2}
Dai Aoki, et al, to be published.

\bibitem{pressure3}
Georg Knebel, Motoi Kimata, Michal Vali\v{s}ka, Fuminori Honda, DeXin Li, Daniel Braithwaite, G\'{e}rard Lapertot, William Knafo, Alexandre Pourret, Yoshiki J. Sato, Yusei Shimizu, Takumi Kihara, Jean-Pascal Brison, Jacques Flouquet, and Dai Aoki,
Anisotropy of the Upper Critical Field
in the Heavy-Fermion Superconductor UTe2 under Pressure,
J. Phys. Soc. Jpn. {\bf 89}, 053707 (2020).

\bibitem{pressure4}
Dai Aoki, Motoi Kimata, Yoshiki J. Sato, Georg Knebel, Fuminori Honda, Ai Nakamura, Dexin Li, Yoshiya Homma, Yusei Shimizu, William Knafo, Daniel Braithwaite, Michal Vali\v{s}ka, Alexandre Pourret, Jean-Pascal Brison, and Jacques Flouquet,
Field-Induced Superconductivity near the Superconducting Critical Pressure in UTe$_2$,
J. Phys. Soc. Jpn. {\bf 90}, 074705 (2021).

\bibitem{butch1}
Wen-Chen Lin, Daniel J. Campbell, Sheng Ran, I-Lin Liu, Hyunsoo Kim, Andriy H. Nevidomskyy, David Graf, Nicholas P. Butch, 
and Johnpierre Paglione,
Tuning magnetic confinement of spin-triplet superconductivity,
npj Quantum Mater. {\bf 5}, 68 (2020).


\bibitem{butch2}
Sheng Ran, Shanta R. Saha, I-Lin Liu, David Graf, Johnpierre Paglione, and Nicholas P. Butch,
Expansion of the high field-boosted superconductivity in UTe$_2$ under pressure,
npj Quantum Mater. {\bf 6}, 75 (2021).


\bibitem{butch3}
Hyunsoo Kim, I-Lin Liu, Wen-Chen Lin, Yun Suk Eo, Sheng Ran, Nicholas P. Butch, and Johnpierre Paglione,
Tuning a magnetic energy scale with pressure in UTe$_2$,
arXiv:2307.00180.

\bibitem{shimizu4}

Yusei Shimizu, Shunichiro Kittaka, Yohei Kono, Toshiro Sakakibara, Kazushige Machida, Ai Nakamura, Dexin Li, Yoshiya Homma, Yoshiki J. Sato, Atsushi Miyake, Minoru Yamashita, and Dai Aoki,
Anomalous electromagnetic response in the spin-triplet superconductor 
UTe$_2$,
Phys. Rev. B {\bf 106}, 214525 (2022).

\bibitem{kaptulnik}
Di S. Wei, David Saykin, Oliver Y. Miller, Sheng Ran, Shanta R. Saha, 
Daniel F. Agterberg, J{\"o}rg Schmalian, Nicholas P. Butch, Johnpierre Paglione, and Aharon Kapitulnik,
The interplay between magnetism and superconductivity in UTe$_2$,
Phys. Rev. B {\bf 105},  024521 (2022).


\bibitem{madhavan}
Lin Jiao, Zhenyu Wang, Sheng Ran, Jorge Olivares Rodriguez, Manfred Sigrist, Ziqiang Wang, 
Nicholas Butch, and Vidya Madhavan,
Microscopic evidence for a chiral superconducting order parameter in the heavy fermion superconductor UTe$_2$,
Nature, {\bf 579}, 523 (2020).




\bibitem{metz} 
T. Metz, S. Bao, S. Ran, I-L. Liu, Y. S. Eo, and W. T. Fuhrman, 
D. F. Agterberg, S. Anlage, N. P. Butch, and J. Paglione, 
Point node gap structure of spin-triplet superconductor UTe$_2$,
Phys. Rev. B {\bf 100}, 220504 (R) (2019).


\bibitem{kittaka}
Shunichiro Kittaka, Yusei Shimizu, Toshiro Sakakibara, Ai Nakamura, Dexin
Li, Yoshiya Homma, Fuminori Honda, Dai Aoki, and Kazushige Machida, 
Orientation of point nodes 
and nonunitary triplet pairing tuned by the easy-axis magnetization in UTe$_2$,
Phys. Rev. Research {\bf 2}, 032014(R) (2020).



\bibitem{sonier1}
N. Azari, M. Yakovlev, N. Rye, S. R. Dunsiger, S. Sundar, M. M. Bordelon, 
S. M. Thomas, J. D. Thompson, P. F. S. Rosa, and J. E. Sonier,
Absence of spontaneous magnetic fields due to time-reversal symmetry 
breaking in bulk superconducting UTe$_2$,
to be published in PRL.


\bibitem{shibauchi}
K. Ishihara, M. Roppongi, M. Kobayashi, Y. Mizukami, H. Sakai, Y. Haga, K. Hashimoto, and  T. Shibauchi,
Chiral superconductivity in UTe$_2$ probed by anisotropic low-energy excitations,
Nat. Commun. {\bf 14}, 2966 (2023)


\bibitem{matsuda}
S. Suetsugu, M. Shimomura, M. Kamimura, T. Asaba, H. Asaeda, Y. Kosuge, Y. Sekino, S. Ikemori, Y. Kasahara, Y. Kohsaka, M. Lee, Y. Yanase, H. Sakai, P. Opletal, Y. Tokiwa, Y. Haga, and Y . Matsuda,
Fully gapped pairing state in spin-triplet superconductor UTe$_2$,
arXiv:2306.17549.

\bibitem{machida0} 
K. Machida and T. Ohmi, 
Phenomenological theory of ferromagnetic superconductivity,
Phys. Rev. Lett. {\bf 86}, 850 (2001).



\bibitem{annett} 
J. F. Annett,
Symmetry of the order parameter for high-temperature superconductivity,
Adv.  Phys. {\bf 39}, 83 (1990).


\bibitem{ozaki1} 
Masa-aki Ozaki, Kazushige Machida, and Tetsuo Ohmi,
On p-Wave Pairing Superconductivity under Cubic Symmetry, 
Prog. Theor. Phys. {\bf 74}, 221 (1985).

\bibitem{ozaki2} 
Masa-aki Ozaki, Kazushige Machida, and Tetsuo Ohmi,
On p-Wave Pairing Superconductivity under Hexagonal and Tetragonal Symmetries, 
Prog. Theor. Phys. {\bf 75}, 442 (1986).



\bibitem{ramires}
Aline Ramires,
Nonunitary Superconductivity in Complex Quantum Materials,
J. Phys.: Condens. Matter {\bf 34}, 304001 (2022).


\bibitem{sonier}
S. Sundar, S. Gheidi, K. Akintola, A. M. C\^{o}t\`{e}, S. R. Dunsiger, 
S. Ran, N. P. Butch, S. R. Saha, J. Paglione, and J. E. Sonier,
Coexistence of ferromagnetic fluctuations and superconductivity in the actinide superconductor UTe$_2$,
Phys. Rev. B {\bf 100}, 140502 (R) (2019).


\bibitem{tokunaga1}
Yo Tokunaga, Hironori Sakai, Shinsaku Kambe, Taisuke Hattori, Nonoka Higa, Genki Nakamine, Shunsaku Kitagawa, 
Kenji Ishida, Ai Nakamura, Yusei Shimizu, Yoshiya Homma, DeXin Li, Fuminori Honda, and Dai Aoki,
$^{125}$Te-NMR study on a single crystal of heavy fermion superconductor UTe$_2$,
 J. Phys. Soc. Jpn. {\bf 88}, 073701 (2019).
 
 \bibitem{tokunaga2}
Yo Tokunaga, Hironori Sakai, Shinsaku Kambe, Yoshinori Haga, Yoshifumi Tokiwa, 
Petr Opletal, Hiroki Fujibayashi, Katsuki Kinjo, Shunsaku Kitagawa, Kenji Ishida, Ai Nakamura, 
Yusei Shimizu, Yoshiya Homma, Dexin Li, Fuminori Honda, and Dai Aoki,
Slow Electronic Dynamics in the Paramagnetic State of UTe$_2$,
 J. Phys. Soc. Jpn. {\bf 91}, 023707 (2022).


\bibitem{furukawa}
 Devi V. Ambika, Qing-Ping Ding, Khusboo Rana, Corey E. Frank, Elizabeth L. Green, 
 Sheng Ran, Nicholas P. Butch, and Yuji Furukawa,
Possible Coexistence of Antiferromagnetic and Ferromagnetic Spin Fluctuations in the Spin-triplet Superconductor 
UTe$_2$ Revealed by $^{125}$Te NMR under Pressure,
 Phys. Rev. B {\bf 105}, L220403 (2022).
 

\bibitem{tokunaga-prl}
Y. Tokunaga, H. Sakai, S. Kambe, P. Opletal, Y. Tokiwa, Y. Haga, S. Kitagawa, K. Ishida, 
D. Aoki, G. Knebel, G. Lapertot, S. Kr\"{a}mer, and M. Horvati\'{c},
Longitudinal spin fluctuations driving field-reinforced superconductivity in UTe$_2$,
Phys. Rev. Lett. {\bf 131}, 226503 (2023).


\bibitem{aeppli}
G. Aeppli, E. Bucher, C. Broholm, J. K. Kjems, J. Baumann, and J. Hufnagl,
Magnetic order and fluctuations in superconducting UPt$_3$,
Phys. Rev. Lett. {\bf 60}, 615  (1988).

\bibitem{bandfolding}
D. Song, S. Lee, Z. Shen, W. Jung, W. Lee, S. Choi, W. Kyung, S. Jung, S. Ishida, S-R. Park, 
H. Eisaki, Y. Wang, K-Y. Choi, and C. Kim,
Interplay between hole superconductivity and quantum critical antiferromagnetic fluctuations in 
electron-doped cuprates,
preprint DOI:10.21203/rs.3.rs-3333329/v1.



\bibitem{anderson}
P. W. Anderson,
Structure of ``triplet" superconducting energy gaps,
Phys. Rev. B {\bf 30}, 4000 (1984).

\bibitem{blount}
E. I. Blount,
Symmetry properties of triplet superconductors,
Phys. Rev. B {\bf 32}, 2935 (1985).

\bibitem{gorkov}
G. E. Volovik, and L. P. Gor'kov,  
Superconducting classes in heavy-fermion systems,
Sov. Phys. JETP {\bf 61}, 843 (1985).





\bibitem{mermin}
V. Ambegaokar and N. D. Mermin, 
Thermal anomalies of  He$^3$:  pairing in a magnetic field,
Phys. Rev. Lett. {\bf 30}, 81 (1973).



\bibitem{repulsion1} 
K. Machida, T. Ohmi, and M. Ozaki,
Anisotropy of Upper Critical Fields for d- and p-Wave Pairing Superconductivity,
J. Phys. Soc. Jpn. {\bf 54}, 1552 (1985).

\bibitem{repulsion2} 
K. Machida, M. Ozaki, and T. Ohmi,
Unconventional Superconducting Class in a Heavy Fermion System UPt$_3$,
J. Phys. Soc. Jpn. {\bf 59}, 1397 (1990).

\bibitem{repulsion3} 
K. Machida, T. Fujita, and T. Ohmi,
Vortex Structures in an Anisotropic Pairing Superconducting State with Odd-Parity,
J. Phys. Soc. Jpn. {\bf 62}, 680 (1993).

\bibitem{repulsion4} 
K. Machida, T. Nishira, and T. Ohmi,
Orbital Symmetry of a Triplet Pairing in a Heavy Fermion Superconductor UPt$_3$,
J. Phys. Soc. Jpn. {\bf 68}, 3364 (1999).


\bibitem{tinkham}
M. Tinkham, {\it Introduction to Superconductivity,} 
McGraw-Hill, New York, 1975.


\bibitem{jp}
V. Jaccarino and M. Peter,
Ultra-High-Field Superconductivity,
Phys. Rev. Lett. {\bf 9}, 290 (1962).



\bibitem{miyake} 
A. Miyake, Y. Shimizu, Y. J. Sato, D. Li, A. Nakamura, Y. Homma, F. Honda, 
J. Flouquet, M. Tokunaga, and D. Aoki, 
Metamagnetic transition in heavy Fermion superconductor UTe$_2$,
 J. Phys. Soc. Jpn. {\bf 88}, 063706 (2019).
 
\bibitem{initial} 
D. Aoki, private communication.


\bibitem{miyake2} 
Atsushi Miyake, Yusei Shimizu, Yoshiki J. Sato, Dexin Li, Ai Nakamura, Yoshiya Homma, 
Fuminori Honda, Jacques Flouquet, Masashi Tokunaga, and Dai Aoki,
Enhancement and Discontinuity of Effective Mass through the First-Order 
Metamagnetic Transition in UTe$_2$,
 J. Phys. Soc. Jpn. {\bf 90}, 103702 (2021).



\bibitem{takagi}
S. Takagi,
Susceptibility Discontinuity at the He$^3$-A-: Normal Transition,
Prog. Theor. Phys. {\bf 51}, 1998 (1974).

\bibitem{aokidHvA}
D. Aoki, H. Sakai, P. Opletal, Y. Tokiwa, J. Ishizuka, Y. Yanase, H. Harima, 
A. Nakamura, D. Li, Y. Homma, Y. Shimizu, G. Knebel, J. Flouquet, and Y. Haga,
First Observation of the de Haas-van  Alphen Effect and 
Fermi Surfaces in the Unconventional Superconductor UTe$_2$,
J. Phys. Soc, Jpn. {\bf 91},  083704 (2022).



\bibitem{miyakeriron}
K. Miyake,
Theory of Pairing Assisted Spin Polarization in Spin-Triplet Equal Spin Pairing: Origin 
of Extra Magnetization in Sr$_2$RuO$_4$ in Superconducting State,
J. Phys. Soc, Jpn. {\bf 83},  053701 (2014).


\bibitem{tsutsumi}
Y. Tsutsumi and K. Machida,
Topological spin texture and d-vector rotation in spin-triplet superconductors: A case of UTe$_2$,
arXiv:2309.02918.

\bibitem{knafo2}
W. Knafo, T. Thebault, P. Manuel, D. D. Khalyavin, F. Orlandi, E. Ressouche, 
K. Beauvois, G. Lapertot, K. Kaneko, D. Aoki, D. Braithwaite, G. Knebel, and S. Raymond,
Incommensurate antiferromagnetism in UTe$_2$ under pressure,
arXiv:2311.05455.

\bibitem{roman} 
Sangyun Lee, Andrew J. Woods, P. F. S. Rosa, S. M. Thomas, E. D. Bauer, Shi-Zeng Lin, and R. Movshovich,
Anisotropic field-induced changes in the superconducting order parameter of UTe$_2$,
arXiv:2310.04938.


\bibitem{matsumura2}
H. Matsumura, et al, presented at the Autumn Meeting of the Japan Physical Society. September 2023.

\bibitem{shibauchi2}
K. Ishihara, M. Kobayashi, K. Imamura, M. Konczykowski, H. Sakai, P. Opletal, 
Y. Tokiwa, Y. Haga, K. Hashimoto, and T. Shibauchi,
Anisotropic Enhancement of Lower Critical Field in 
Ultraclean Crystals of Spin-Triplet Superconductor UTe$_2$,
Phys. Rev. Research {\bf 5}, L022002 (2023).

\bibitem{paulsen}
C. Paulsen, G. Knebel, G. Lapertot, D. Braithwaite, A. Pourret, D. Aoki, F. Hardy, J. Flouquet, and J.-P. Brison,
Anomalous anisotropy of the lower critical field and Meissner effect in UTe$_2$,
Phys. Rev. B {\bf 103}, L180501 (2021).


\bibitem{isoshima}
T. Isoshima and K. Machida, 
Instability of the nonvortex toward a quantized vortex in a  Bose-Einstein condensate
under rotation,
Phys. Rev. A {\bf 60}, 3313 (1999).


 \bibitem{matsu}
 K. Machida,
 Spin Density Wave and Superconductivity in Highly Anisotropic Materials,
J. Phys. Soc. Jpn. {\bf 50}, 2195 (1981). 
 K. Machida and T. Matsubara,
 Spin Density Wave and Superconductivity in Highly Anisotropic Materials. II. 
 Detailed Study of Phase Transitions,
 J. Phys. Soc. Jpn. {\bf 50}, 3231 (1981).

 \bibitem{nokura}
 K. Machida, K. Nokura, and T. Matsubara,
 Theory of antiferromagnetic superconductors,
Phys. Rev. B {\bf 22}, 2307 (1980).
 
 \bibitem{nakanishi}
 K. Machida and H. Nakanishi,
Superconductivity under a ferromagnetic molecular field,
Phys. Rev. B {\bf 30}, 122 (1984).
 
 \bibitem{kato}
 K. Machida and M. Kato,
Inherent Spin-Density-Wave Instability in Heavy-Fermion Superconductivity, 
 Phys. Rev. Lett. {\bf 58}, 1986 (1987).
 
 
\bibitem{kivelson}
Eduardo Fradkin, Steven A. Kivelson, and John M. Tranquada,
Theory of intertwined orders in high temperature superconductors,
Rev. Mod. Phys. {\bf 87}, 457 (2015).

\bibitem{keimer}
B. Keimer, S. A. Kivelson, M. R. Norman, S. Uchida, and J. Zaanen,
From quantum matter to
high-temperature superconductivity in copper oxides,
Nature {\bf 518}, 179 (2015).




 \bibitem{ichioka}
 M. Ichioka, N. Hayashi, and K. Machida,
 Local density of states in the vortex lattice in a type-II superconductor,
Phys. Rev. B {\bf 55}, 6565 (1997).


\bibitem{andersen}
Henrik S. R{\o}ising, Max Geier, Andreas Kreisel, and Brian M. Andersen,
Thermodynamic transitions and topology of spin-triplet superconductivity: Application to UTe$_2$,
arXiv:2311.06097.




\bibitem{miranovic}
P. Miranovi\'c, N. Nakai, M. Ichioka, and K. Machida,
Orientational field dependence of low-lying excitations in the mixed state
of unconventional superconductors,
Phys. Rev. B {\bf 68}, 052501 (2003).


\end{thebibliography}
\end{document}